\documentstyle[12pt,titlepage]{article}
\newcommand{\be}{\begin{equation}}
\newcommand{\ee}{\end{equation}}
\newcommand{\bea}{\begin{eqnarray}}
\newcommand{\eea}{\end{eqnarray}}
\newcommand{\ket}{\rangle}
\newcommand{\bra}{\langle}
\newcommand{\bm}[1]{\mbox{\bf #1}}

\newcounter{SG2}
\makeatletter
\def\lsim{\mathrel{\mathpalette\subsim@align<}}
\def\gsim{\mathrel{\mathpalette\subsim@align>}}
\def\subsim@align#1#2{\lower.6ex\vbox{\baselineskip\z@skip\lineskip\z@
\ialign{$\m@th#1\hfil##\hfil$\crcr#2\crcr\sim\crcr}}}
\makeatother
\begin{document}

\pagestyle{empty}
\begin{titlepage}

\title{Microscopic description of nuclei\\
in the middle of the pf-shell\\
by a shell model calculation\\
with $G$-matrix interaction}

\author{Hitoshi Nakada\\
{\small Department of Physics, School of Medicine, Juntendo University}\\
{\small Hiraga-gakuendai 1-1, Inba-mura, Inba-gun, Chiba 270-16, Japan}\\
Takashi Sebe\\
{\small Department of Applied Physics, College of Engineering, Hosei
University}\\
{\small Kajino-cho 3-7-2, Koganei, Tokyo 184, Japan}\\
and\\
Takaharu Otsuka\\
{\small Department of Physics, Faculty of Science, University of Tokyo}\\
{\small Hongo 7-3-1, Bunkyo-ku, Tokyo 113, Japan}}

\date{}

\maketitle
\thispagestyle{empty}

%\noindent {\small PACS numbers: 21.60.Cs, 21.60.Ev, 25.30.Dh, 27.40.+z}
%\\

\begin{abstract}
Energy levels and electromagnetic properties of 24 nuclides with $N=28\sim 30$
are studied in terms of a large-scale shell model calculation,
which contains no newly adjusted parameters.
The Kuo-Brown $G$-matrix interaction is shown
to reproduce energy levels of 205 low-lying states of all these nuclei.
We evaluate effective charges by incorporating the core-polarization effects
caused by the coupling to GQR's.
We then compute E2 moments and transition probabilities.
The M1 moments and transition rates are calculated
by quoting the effective $g$-factors of Towner,
which are obtained by taking into account the meson-exchange
and the core-polarization mechanisms.
By this microscopic calculation
most of the E2 properties and the magnetic moments are reproduced.
Although there are agreements and disagreements in the M1 transition rates,
the general tendency is reproduced.
The $(e,e')$ and $(p,p')$ excitation
from the ground state to some low-lying $2^+$ states
is also discussed.
\end{abstract}

\end{titlepage}
\pagestyle{plain}
\setcounter{page}{1}
\setcounter{SG2}{2}

\section{Introduction}
\label{sec:intro}

The nuclear shell model has been a basic tool
to understand the low energy phenomena of nuclei
in terms of the nucleonic degrees of freedom.
Although the shell model has been successful
in the p- and sd-shell regions\cite{ref:sd-shell},
it is prohibitively difficult to apply the model
to most of heavy nuclei,
because of the large size of the model space.
The effective interaction between valence nucleons
should originate in the interaction between free nucleons.
There have been many investigations to derive an effective interaction
from a microscopic viewpoint of this sort.
The restricted model space which enables us
to carry out the shell model calculation
makes it necessary to renormalize the interaction.
The renormalization appears to be successful
in comparison to experiment,
although the convergence of the renormalization procedure
is still being investigated.
Nevertheless, it is known that realistic interactions
derived from free-nucleon interactions
substantially explain the main features of low-lying states
of sd-shell nuclei.
An example is the interaction obtained
in the earlier work of Kuo and Brown\cite{ref:KBsd}.

In the pf-shell, shell model calculations
have been attempted for many years.
In most earlier calculations,
the excitation across the gap at $Z, N=28$,
as well as those at $Z, N=20$ or $40$, was ignored.
Empirical effective interactions were often employed
in those works\cite{ref:McCullen,ref:Vervier,ref:Cohen-Auerbach}.
Horie and Ogawa reported a shell model calculation
for $N=29$ and $30$ nuclei\cite{ref:Ogawa,ref:Ogawa2},
which reproduces yrast levels systematically.
A typical example is $^{56}$Fe.
We, however, cannot reproduce higher states
in the framework of Horie and Ogawa's approach.
For instance, the excitation energy of the $2_3^+$ state of $^{56}$Fe
is overestimated by $0.4$MeV,
and that of the $2_4^+$ state by $0.5$MeV.
This discrepancy is serious in discussing
the mixed-symmetry collective states\cite{ref:Nakada},
which have been attracting a certain amount of
interest\cite{ref:Nakada,ref:Eid,ref:Collins,ref:Lieb,ref:Takamatsu}.

In contrast to the phenomenological interactions,
Kuo and Brown applied their $G$-matrix method to the pf-shell region,
and derived a realistic interaction on the top of
the $^{40}$Ca core\cite{ref:KBpf}.
The Kuo-Brown interaction has been applied to shell model calculations
in the full pf-shell,
for $40<A\leq 44$ nuclei\cite{ref:Motoba2}.
Though qualitative agreement with the experimental data is found
in those calculations,
there still remain recognizable discrepancies.
A few ways of modification of the interaction
have been proposed\cite{ref:Pasquini,ref:Poves,ref:McGrory}.
However, we should take notice
that excitations from the sd-shell may be significantly present
in the mass region of $40<A\leq 44$.
It is worthwhile to test the Kuo-Brown interaction
in a sufficiently large model space
around the middle of the pf-shell,
where low-lying states will minimally be affected by other shells.

Oberlechner and Richert carried out a shell model calculation
around $^{56}$Ni\cite{ref:Oberlechner},
by using another realistic interaction.
They, however, could not succeed
in reproducing the experimental energy levels so well,
presumably because of the truncation of the model space.

There have been several attempts to take into account
the one-particle excitation
from the ${\rm 0f}_{7/2}$
to the ${\rm 1p}_{3/2}$ orbit\cite{ref:Raz-etal}.
There have been shell model calculations
which included the one-particle excitation
from the ${\rm 0f}_{7/2}$
to any of the ${\rm 0f}_{5/2}$, ${\rm 1p}_{3/2}$
and ${\rm 1p}_{1/2}$ orbits\cite{ref:Motoba-etal}.
However, the situation has not necessarily improved
by this extension of the configurations;
the discrepancies with $Ex(2_3^+)$ and $Ex(2_4^+)$ of $^{56}$Fe,
for instance,
remain similar to the results in Ref.\cite{ref:Ogawa2}.
This probably happens
because the excitation of two nucleons
from the ${\rm 0f}_{7/2}$ to the higher orbits is crucial
and can indeed be driven by the deformation
and/or the pairing correlation.

In this paper we show the results of
the large-scale shell model calculation
for $20<Z\leq 28$, $28\leq N\leq 30$ nuclei,
including the excitation of up to two particles (four for $^{56}$Ni)
from the ${\rm 0f}_{7/2}$ to the other pf-shell orbits,
which seems to be of particular importance in this region,
as mentioned just above.
It turns out that,
if a sufficiently large configuration space is chosen,
the Kuo-Brown interaction appears to be quite an excellent
effective interaction.
The model space and the effective hamiltonian are discussed
in more detail in Section~\ref{sec:space&hamil}.
In Section~\ref{sec:energy} calculated energy spectra are shown,
in comparison with the observed ones.
In Sections~\ref{sec:E2} and \ref{sec:M1},
the E2 and M1 properties of those nuclei are analyzed
in terms of the operators obtained microscopically.
Those results will tell us about the reliability of
the shell model wavefunctions,
and also how well we can understand the nuclear properties
from the nucleonic degrees of freedom.
The $(e,e')$ form factors concerning the quadrupole collectivity
are discussed
on the same footing as the E2 transitions, in Section~\ref{sec:FF}.
The DWBA results of $(p,p')$ differential cross sections
are briefly surveyed in Section~\ref{sec:pp'}.
They also supply a justification of the shell model wavefunctions
relevant to the low energy quadrupole collective motion.
Section~\ref{sec:summary} is a summary of this article.

\section{Model space and effective hamiltonian}
\label{sec:space&hamil}

We first show the configuration space of our shell model calculation.
Assuming $^{40}$Ca to be a doubly magic inert core,
we consider configurations such as
\be
 ({\rm 0f}_{7/2})^{n_1-k}
 ({\rm 0f}_{5/2}{\rm 1p}_{3/2}{\rm 1p}_{1/2})^{n_2+k},
  \label{eq:config}
\ee
where $n_1$ and $n_2$ are defined so that $k=0$ should give
the lowest configuration
in which the excitation across the gap
between ${\rm 0f}_{7/2}$ and the other three orbits is absent.
For instance, $n_1=14$ and $n_2=2$ for $^{56}$Fe,
since this nucleus has 6 protons and 10 neutrons in the pf-shell.
For the nuclei handled in this paper, $n_1=(Z-20)+8$, $n_2=N-28$.
The present model space contains all configurations
of $k=0,1$ and $2$ for each nucleus.
We would like to emphasize that such a large configuration space
has never been used except for
Refs.\cite{ref:Nakada,ref:Pasquini,ref:Muto-Horie,ref:Wong-Davies}.
The $k=2$ configurations evidently play an essential role
in the description of the relaxation of the magicity of $Z=N=28$
due to the pairing correlation and the deformation.

Concerning the effective hamiltonian,
we adopt the Kuo-Brown hamiltonian
on top of the $^{40}$Ca core\cite{ref:KBpf}.
In this hamiltonian,
the single-particle energies have been determined from experiment;
$\epsilon_{{\rm 0f}_{7/2}}=0.0$MeV,
$\epsilon_{{\rm 0f}_{5/2}}=6.5$MeV,
$\epsilon_{{\rm 1p}_{3/2}}=2.1$MeV
and $\epsilon_{{\rm 1p}_{1/2}}=3.9$MeV.
The two-body interaction is derived from the $G$-matrix
based on the Hamada-Johnston potential,
assuming the single-particle wavefunction
in the harmonic oscillator approximation.
We include the $3p$-$1h$ correction,
which is evaluated by the perturbation.
It is noted that there are only few reports
of shell model calculations in the pf-shell
using such a microscopic interaction
without any phenomenological modification.
Around the middle of the pf-shell,
both the influence of the sd-shell and that of the sdg-shell
are expected to be minimal.
The Kuo-Brown interaction seems to be valid in this region,
if the configuration space is sufficiently large.
Indeed, the success of the Kuo-Brown interaction
in the $k\leq 2$ configuration space
was reported for $^{48}$Sc in Ref.\cite{ref:Pasquini}.
We already showed the validity also in $^{56}$Fe\cite{ref:Nakada},
and the extensive application is carried out in this study.

The isospin is automatically conserved,
and in the following we will deal with states
with the lowest isospin for each nucleus.

In the numerical calculation, we use an $M$-scheme shell model code
newly developed.
The Lanczos method is applied
for diagonalizing the shell-model hamiltonian\cite{ref:Lanc}.
The present model space is one of the largest one available at present.
In the $M$-scheme, the $M=0$ space has the largest dimension
in even-mass nuclei,
while the $M=\pm {1\over 2}$ space in odd-mass nuclei.
This dimension is displayed for each nucleus in Table~\ref{tab:dim},
as well as the largest dimension in the $JT$-scheme.

\begin{table}
\centering
\caption{\label{tab:dim}}
\begin{small}
The largest dimensions in the shell model calculation
with $k\leq 2$ configurations,
both in the $M$-scheme and the $JT$-scheme.
The fourth and fifth columns indicate the spin-parity and isospin
for which the dimension in the $JT$-scheme is largest.\\
\begin{tabular}{|c|c||r||c|c|r|}
   \hline
     ~$N$~& nucl. & ~$M$-scheme & $J$ & $T$ & $JT$-scheme \\
   \hline
     & $^{49}$Sc & $1,561~$ & ${7\over 2}$, ${9\over 2}$
      & ${7\over 2}$ & $210~$\\
     & $^{50}$Ti & $5,530~$ & $4$ & $3$ & $656~$\\
     & $^{51}$V & $11,676~$ & ${9\over 2}$ & ${5\over 2}$
      & $1,254~$\\
     28 & $^{52}$Cr & $16,544~$ & $5$ & $2$ & $1,615~$\\
     & $^{53}$Mn & $15,883~$ & ${9\over 2}$ & ${3\over 2}$
      & $1,456~$\\
     & $^{54}$Fe & $10,620~$ & $4$ & $1$ & $891~$\\
     & $^{55}$Co & $4,717~$ & ${9\over 2}$ & ${1\over 2}$ & $347~$\\
     & $^{56}$Ni & $1,353~$ & $3$ & $1$ & $103~$\\
   \hline
     & $^{50}$Sc & $5,550~$ & $4$ & $4$ & $740~$\\
     & $^{51}$Ti & $20,425~$ & ${9\over 2}$ & ${7\over 2}$
      & $2,423~$\\
     & $^{52}$V & $46,034~$ & $5$ & $3$ & $4,963~$\\
     29 & $^{53}$Cr & $68,355~$ & ${9\over 2}$ & ${5\over 2}$
      & $6,929~$\\
     & $^{54}$Mn & $70,740~$ & $5$ & $2$ & $6,751~$\\
     & $^{55}$Fe & $50,425~$ & ${9\over 2}$ & ${3\over 2}$
      & $4,611~$\\
     & $^{56}$Co & $24,844~$ & $4$ & $1$ & $2,126~$\\
     & $^{57}$Ni & $7,890~$ & ${7\over 2}$ & ${1\over 2}$ & $619~$\\
   \hline
     & $^{51}$Sc & $13,411~$ & ${9\over 2}$ & ${9\over 2}$
      & $1,774~$\\
     & $^{52}$Ti & $52,624~$ & $4$ & $4$ & $6,207~$\\
     & $^{53}$V & $123,519~$ & ${9\over 2}$ & ${7\over 2}$
      & $13,574~$\\
     30 & $^{54}$Cr & $195,334~$ & $5$ & $3$ & $20,074~$\\
     & $^{55}$Mn & $211,833~$ & ${9\over 2}$ & ${5\over 2}$
      & $20,992~$\\
     & $^{56}$Fe & $162,358~$ & $5$ & $2$ & $15,457~$\\
     & $^{57}$Co & $84,978~$ & ${9\over 2}$ & ${3\over 2}$
      & $8,029~$\\
     & $^{58}$Ni & $29,792~$ & $4$ & $1$ & $2,780~$\\
   \hline
\end{tabular}
\end{small}
\end{table}

We add some remarks on the effective interaction:
the realistic interaction has long been discussed in the sd-shell.
The bare $G$-matrix is insufficient to reproduce
the observed energy spectra,
and some renormalization is required.
Though the $3p$-$1h$ correction improves the energy spectra
significantly\cite{ref:KBsd},
the convergence of the renormalization procedure
has been a serious problem.
Despite much effort and a certain progress\cite{ref:effint},
this problem has not been fully solved yet,
and it is beyond the scope of this article.
We only comment that the $3p$-$1h$ correction gives reasonable spectra
and that the low-lying energy spectra
resulting from an interaction including the higher-order corrections
do not look quite different from
those obtained by the $3p$-$1h$ correction\cite{ref:Shurpin}.

In the $20\leq Z<28$, $20\leq N<28$ region,
there are many intruder states,
which are not described within the shell model for the pf-shell.
These intruder states will primarily consist of
the excitation from the sd-shell orbits to ${\rm 0f}_{7/2}$.
Whereas one hole in the sd-shell leads to negative-parity states,
two-hole configuration could be low in energy owing to collectivity.
The proton-neutron interaction should play a dominant role
in introducing this collectivity.
The coexistence of the intruder configuration significantly
influences the spectrum,
and cannot be treated within the renormalization scheme
of Kuo-Brown's $3p$-$1h$ correction,
which is based on the perturbation theory.
The modification of the Kuo-Brown interaction discussed
in Refs.\cite{ref:Pasquini,ref:Poves}
yields a better agreement in the $20\leq Z<28$, $20\leq N<28$ region,
and this may be connected with the influence
of the intruder configuration.
The influence of the intruder configuration becomes weaker
in $N\geq 28$ nuclei,
due to the suppression of the neutron excitation
from the sd-shell to ${\rm 0f}_{7/2}$
by the Pauli blocking.
This will also reduce the proton excitation from the $^{40}$Ca core
through the proton-neutron correlation.
On the other hand, as we approach $Z=N=28$,
the excitation from ${\rm 0f}_{7/2}$
to $({\rm 0f}_{5/2}{\rm 1p}_{3/2}{\rm 1p}_{1/2})$ will grow.
In such cases, the extended configuration space currently considered
becomes crucial,
as can be seen in the following sections.

In Ref.\cite{ref:KBpf}, Kuo and Brown tabulated the interaction
for the model space consisting of
the four pf-shell orbits and ${\rm 0g}_{9/2}$.
In this study we omit the contribution
containing the ${\rm 0g}_{9/2}$ orbit,
because the influence of ${\rm 0g}_{9/2}$
is not expected to be significant for the nuclei under consideration.
We will return to this problem of the model space
in Sections~\ref{subsec:Ni57} and \ref{sec:FF}.

We shall use a mass-number independent interaction;
the harmonic oscillator bases have been employed
in calculating the $G$-matrix,
with $\hbar\omega = 10.5$MeV in Ref.\cite{ref:KBpf}.
The oscillator length assumed usually as $b\propto A^{1/6}$
varies by only a few percent among the nuclei under consideration,
namely $A=49\sim 58$.
The above value $\hbar\omega = 10.5$MeV adopted in Ref.\cite{ref:KBpf}
corresponds to $A\simeq 60$ if we take $\hbar\omega = 41.2 A^{-1/3}$.
Thus, the $G$-matrix calculated in Ref.\cite{ref:KBpf}
appears to be appropriate for the mass region considered in this study.

\section{Energy levels}
\label{sec:energy}

In this Section we systematically show results of
the present calculation
for energy levels of $20<Z\leq 28$, $28\leq N\leq 30$ nuclei,
in comparison with experimental data.
The data are quoted from the compilation
in Refs.\cite{ref:NDS49}--\cite{ref:NDS58}.
We concentrate on the excitation energies and
do not discuss the binding energies.

\subsection{Ground-state configuration and convergence by truncation}
\label{subsec:config}

Before going to the energy levels,
we shall see the ground-state configurations
classified in terms of $k$ in Eq.(\ref{eq:config}).
They are shown in Table~\ref{tab:config}.

\begin{table}
\centering
\caption{\label{tab:config}}
\begin{small}
Ground-state configuration in the shell model calculation
in terms of the probabilities ($\%$) of each $k$ configuration.\\
\begin{tabular}{|c|c|c||r|r|r|}
   \hline
     ~$N$~& nucl. & state & $k=0~~$ & $k=1~~$ & $k=2~~$ \\
   \hline
     & $^{49}$Sc & $[{7\over 2}]_1^-$ & $68.2$~~~& $10.3$~~~
     & $21.5$~~~\\
     & $^{50}$Ti & $0_1^+$ & $61.1$~~~& $14.9$~~~& $24.0$~~~\\
     & $^{51}$V & $[{7\over 2}]_1^-$ & $60.5$~~~& $14.7$~~~
     & $24.8$~~~\\
     28 & $^{52}$Cr & $0_1^+$ & $60.8$~~~& $12.3$~~~& $26.9$~~~\\
     & $^{53}$Mn & $[{7\over 2}]_1^-$ & $61.3$~~~& $10.8$~~~
     & $27.9$~~~\\
     & $^{54}$Fe & $0_1^+$ & $62.2$~~~& $ 7.0$~~~& $30.8$~~~\\
     & $^{55}$Co & $[{7\over 2}]_1^-$ & $63.0$~~~& $ 4.5$~~~
     & $32.5$~~~\\
     & $^{56}$Ni & $0_1^+$ & $63.4$~~~& $ 0.0$~~~& $36.6$~~~\\
   \hline
     & $^{50}$Sc & $5_1^+$ & $68.8$~~~& $14.6$~~~& $16.6$~~~\\
     & $^{51}$Ti & $[{3\over 2}]_1^-$ & $60.7$~~~& $19.2$~~~
     & $20.1$~~~\\
     & $^{52}$V & $3_1^+$ & $56.6$~~~& $22.9$~~~& $20.5$~~~\\
     29 & $^{53}$Cr & $[{3\over 2}]_1^-$ & $57.3$~~~& $20.2$~~~
     & $22.5$~~~\\
     & $^{54}$Mn & $3_1^+$ & $55.5$~~~& $21.6$~~~& $22.9$~~~\\
     & $^{55}$Fe & $[{3\over 2}]_1^-$ & $58.4$~~~& $15.1$~~~
     & $26.5$~~~\\
     & $^{56}$Co & $4_1^+$ & $59.4$~~~& $12.1$~~~& $28.5$~~~\\
     & $^{57}$Ni & $[{3\over 2}]_1^-$ & $60.9$~~~& $ 6.3$~~~
     & $32.8$~~~\\
   \hline
     & $^{51}$Sc & $[{7\over 2}]_1^-$ & $70.2$~~~& $10.8$~~~
     & $19.0$~~~\\
     & $^{52}$Ti & $0_1^+$ & $62.7$~~~& $14.8$~~~& $22.5$~~~\\
     & $^{53}$V & $[{7\over 2}]_1^-$ & $59.8$~~~& $19.3$~~~
     & $20.8$~~~\\
     30 & $^{54}$Cr & $~0_1^+~$ & $56.5$~~~& $20.6$~~~& $22.9$~~~\\
     & $^{55}$Mn & $[{5\over 2}]_1^-$ & $53.2$~~~& $25.1$~~~
     & $21.7$~~~\\
     & $^{56}$Fe & $0_1^+$ & $55.7$~~~& $17.7$~~~& $26.6$~~~\\
     & $^{57}$Co & $[{7\over 2}]_1^-$ & $56.8$~~~& $15.4$~~~
     & $27.9$~~~\\
     & $^{58}$Ni & $0_1^+$ & $57.7$~~~& $ 9.1$~~~& $33.1$~~~\\
   \hline
\end{tabular}
\end{small}
\end{table}

Though one might be interested in the Ca isotopes,
they are not suitable for the present study.
Probably, in the Ca isotopes, we need to take into account
the excitation from the sd-shell.

Within the present calculation,
it turns out that the ground state of any nucleus is dominated
by the $k=0$ configuration, as is expected.
The ground state of $^{56}$Ni is
outside the $({\rm 0f}_{7/2})^{16}$
closed-core by $37\%$.
The $k=0$ probabilities are around $60\%$ except for Sc isotopes,
while about $70\%$ in Sc isotopes.
Thus, the ground state is dominated by the $k=0$ configuration
in all cases studied.
The importance of the $k=2$ configuration is obvious
in Table~\ref{tab:config};
The probability of the $k=2$ configuration is even larger
than that of the $k=1$ configuration in many of the nuclei
under study.

Table~\ref{tab:config} shows that, for $N=28$, $29$ and $30$,
the $k=2$ probability goes up as $Z$ increases.
This observation gives us information
whether the wavefunction is converging under the $k\leq 2$ truncation.
This convergence will be referred to as $k$-convergence.
As $Z$ is larger, the proton Fermi energy should be higher.
Therefore the valence proton is more susceptible
to be excited from ${\rm 0f}_{7/2}$ to the upper orbits.
This proton excitation will induce the excitation of neutrons
from ${\rm 0f}_{7/2}$
through the proton-neutron correlation.
In other words,
the energy gap above the ${\rm 0f}_{7/2}$ orbit
becomes less important.
Consequently, as we approach $Z=28$,
the $k$-convergence would become relatively worse.
The sd-shell configurations, on the other hand,
could have a stronger influence for $Z$ close to 20,
since the Pauli blocking suppresses to a lesser extent
the proton excitation
from the sd-shell to ${\rm 0f}_{7/2}$.
These problems seem to be much less serious in the region $Z\sim 24$.

The $k=2$ probability is slightly smaller in $N=29$ and $30$
than in $N=28$ isotones.
This happens probably because, in $N=29$ and $30$ isotones,
the partial occupancy of the neutron
$({\rm 0f}_{5/2}{\rm 1p}_{3/2}{\rm 1p}_{1/2})$ orbits
suppresses the excitation from ${\rm 0f}_{7/2}$.
Taking into account the diminishing excitations from the sd-shell,
the $k$-convergence is expected to be quite good
for $N=29$ and $30$ nuclei.

According to the above considerations,
we anticipate that, among the nuclei under discussion,
the $k$-convergence will be worst for $^{56}$Ni.
On the other hand, it is possible to diagonalize
the shell model hamiltonian for $^{56}$Ni
in the space containing all the $k\leq 4$ configurations,
which leads to the $M$-scheme dimension of $497,805$.
By this calculation, the ground-state energy of $^{56}$Ni
is lowered by 3.3MeV,
and the $k=0$ probability reduces to $28\%$.
We shall return to this $k\leq 4$ calculation
in Subsection~\ref{subsec:Ni56-Co55}.

While there seems to be a slight difference
in the $k$-convergence of the wavefunctions,
the energy levels are excellently reproduced,
as is discussed in detail in the subsequent subsections.
We will discuss the convergence
also in Sections~\ref{sec:E2} and \ref{subsec:sum_EM}.

\subsection{$^{56}$Ni, $^{49}$Sc and $^{55}$Co}
\label{subsec:Ni56-Co55}

In the $k=0$ configuration space,
we have a single state with $J^P=0^+$ in $^{56}$Ni,
and one with $J^P={7\over 2}^-$ in $^{49}$Sc and $^{55}$Co.
These spin-parities of the ground states are unchanged
in the present calculation, in agreement with experiments.

The first excited levels become too high in $^{56}$Ni and $^{55}$Co
in the $k\leq 2$ calculation.
In order to describe these states,
one probably has to consider a softer $^{56}$Ni core
than the present one.
Higher $k$ configurations will be necessary
in order to reproduce those levels.

Wong and Davies carried out a shell model calculation for $^{56}$Ni
by including the $k\leq 4$ configuration in part\cite{ref:Wong-Davies}.
They reported that the Kuo-Brown interaction successfully reproduces
the energy spectrum of $^{56}$Ni,
although their single particle energies are somewhat different
from the present ones.
There the $({\rm 0f}_{7/2})^{16}$ probability of
the ground-state wavefunction was $16\%$.
The small $({\rm 0f}_{7/2})^{16}$ probability
has brought about
several discussions\cite{ref:Pasquini,ref:Oberlechner}.
In the $k\leq 4$ calculation for $^{56}$Ni
with the present hamiltonian,
the ground state seems to have a similar structure to
the result of Wong and Davies,
though the $k=0$ probability is larger in the present result.
In Table~\ref{tab:Ni56},
the excitation energy of the $2_1^+$ level
obtained by the $k\leq 4$ calculation,
as well as by the $k\leq 2$ calculation,
are shown in comparison with the datum.
The $k\leq 2$ calculation yields too high an excitation energy,
suggesting that the $k\leq 2$ model space is too small for this state.
On the other hand, the observed energy is reproduced quite well
by the $k\leq 4$ calculation.
This agreement encourages us in applying
the present $G$-matrix interaction
to the nuclei around $^{56}$Ni.
The $k\leq 4$ calculation yields a $0^+$ level at $Ex=2.30$MeV,
which is strongly dominated by the $k=4$ configuration.
An unobserved $2^+$ level also appears at $Ex=2.96$MeV
in this calculation.
It is interesting to know whether or not these levels exist.
The low-lying levels of $^{56}$Ni obtained
by the present $k\leq 4$ calculation
are summarized in Fig.\ref{fig:enNi56x},
in comparison with the data\cite{ref:NDS56}.
We note that the indispensable roles of
$k=3$ and $4$ configurations in $^{56}$Ni
are due to $Z=N=28$ and also due to
that neither excited $0^+$ nor $2^+$ is allowed with $k=0$.

\begin{table}
\centering
\caption{\label{tab:Ni56}}
\begin{small}
Comparison of the $k\leq 2$ and $k\leq 4$ calculation
with the experimental data taken from Refs.\cite{ref:NDS56},
for $Ex(2_1^+)$ (MeV) and
$B({\rm E2}; 2_1^+\rightarrow 0_1^+)$ values ($e^2 {\rm fm}^4$)
in $^{56}$Ni.\\
\begin{tabular}{|cc||c|c||c|}
   \hline
     \multicolumn{2}{|c||}{quantity} & Cal. ($k\leq 2$)
      & Cal. ($k\leq 4$) &~~~~~Exp.~~~~~ \\
   \hline
     $Ex(2_1^+)$ &[MeV]& $5.792$ & $2.724$ & $2.701$ \\
     $B({\rm E2}; 2_1^+\rightarrow 0_1^+)$ &[$e^2 {\rm fm}^4$]
     & $80.4$ & $80.6$ & $74~\pm~^{36}_{28}$ \\
   \hline
\end{tabular}
\end{small}
\end{table}

\begin{figure}
\centering
%\vspace{80mm}
\vspace{30mm}
\caption{\label{fig:enNi56x}}
\begin{small}
Experimental and calculated energy levels of $^{56}$Ni.
The experimental data are taken from Ref.\cite{ref:NDS57}.
The calculated energy levels are obtained
from the present $k\leq 4$ shell model calculation.
\end{small}
\end{figure}

In $^{49}$Sc, no negative-parity states have been observed
below $Ex=3$MeV
other than the ground state.
Nevertheless there emerge about ten negative-parity states
in $2.4<Ex<3$MeV in the calculation.
Since $^{49}$Sc is far from $\beta$-stable line,
those states may not have been observed yet.
Because $Z\sim 20$,
the influence of the excitation from the sd-shell
may be sizable in comparison with the larger $Z$ case.

\subsection{$^{57}$Ni}
\label{subsec:Ni57}

Suppose that $^{56}$Ni is taken to be an inert core,
there is only one valence neutron in the $^{57}$Ni nucleus.
Therefore we have only three states with
$J^P={1\over 2}^-$, ${3\over 2}^-$, ${5\over 2}^-$
in the $k=0$ configuration.

In Fig.\ref{fig:enNi57}, the calculated and measured energy levels
are compared.
The agreement is sufficiently good for the lowest three states,
which predominantly consist of the $k=0$ configuration.
On the other hand,
the energy gap between the lowest three states and the other states
is larger in the calculation than in the experiment,
similarly to $^{56}$Ni and $^{55}$Co.
A larger model space will be necessary for these higher states.

\begin{figure}
\centering
%\vspace{80mm}
\vspace{30mm}
\caption{\label{fig:enNi57}}
\begin{small}
Energy levels of $^{57}$Ni.
The experimental data are taken from Ref.\cite{ref:NDS57}.
The calculated energy levels are obtained
from the present $k\leq 2$ shell model calculation.
\end{small}
\end{figure}

In the $k=0$ space with the same hamiltonian,
we have too low an energy
for the $[{5\over 2}]^-_1$ state ($Ex=0.28$MeV).
The $k=1$ and $2$ configurations improves the energy
by raising the $[{5\over 2}]^-_1$ relative
to the $[{3\over 2}]^-_1$ and $[{1\over 2}]^-_1$ states.

Note that no $J^P={9\over 2}^+$ state is observed as low
as the $[{3\over 2}]^-_1$, $[{5\over 2}]^-_1$
and $[{1\over 2}]^-_1$ levels.
A candidate of $[{9\over 2}]^+_1$ is reported experimentally
around $Ex=3$MeV at lowest\cite{ref:NDS57}.
This fact suggests that we do not have to
include the ${\rm 0g}_{9/2}$ orbit explicitly,
as far as we restrict ourselves to the states below about 3MeV.

Recently phenomenological interactions in the pf-shell,
named FPD6 and FPMI3, have been proposed\cite{ref:FPD6&FPMI3}.
These interactions are adjusted for the full pf-shell calculation
in $40<A\leq 44$ nuclei and Ca isotopes.
It is worthwhile testing these interactions in the present mass region.

In the $k\leq 2$ space, neither FPD6 nor FPMI3
reproduces the $^{57}$Ni spectrum;
FPD6 makes the $[{5\over 2}]^-_1$ lower
than the $[{3\over 2}]^-_1$ by 0.5MeV,
while in the case of FPMI3,
both the $[{5\over 2}]^-_1$ and the $[{1\over 2}]^-_1$
become lower than the $[{3\over 2}]^-_1$
by 2.5 and 1.4MeV, respectively.
Note that the probabilities of the $k=0$ configuration are larger
than in the case of the Kuo-Brown interaction
for those three lowest-lying states.
Thus, those FPD6 and FPMI3 results are less sensitive
to higher $k$ configurations.
Both FPD6 and FPMI3 are work better
for lighter pf-shell nuclei\cite{ref:FPD6&FPMI3}.

\subsection{$Z={\em even},~N=28$ nuclei}
\label{subsec:Zeven-N28}

The energy levels in $^{50}$Ti, $^{52}$Cr and $^{54}$Fe
are displayed in Figs.\ref{fig:enTi50}, \ref{fig:enCr52}
and \ref{fig:enFe54}.
We find that the low-lying states correspond well
to the observed levels,
in the energy range of $Ex\lsim 4$MeV,
and with deviations of $\delta E\lsim 0.3$MeV.

\begin{figure}
\centering
%\vspace{100mm}
\vspace{30mm}
\caption{\label{fig:enTi50}}
\begin{small}
Energy levels of $^{50}$Ti.
The data are taken from Ref.\cite{ref:NDS50}.
\end{small}
\end{figure}

\begin{figure}
\centering
%\vspace{100mm}
\vspace{30mm}
\caption{\label{fig:enCr52}}
\begin{small}
Energy levels of $^{52}$Cr.
The data are taken from Ref.\cite{ref:NDS52}.
\end{small}
\end{figure}

\begin{figure}
\centering
%\vspace{100mm}
\vspace{30mm}
\caption{\label{fig:enFe54}}
\begin{small}
Energy levels of $^{54}$Fe.
The data are taken from Ref.\cite{ref:NDS54}.
\end{small}
\end{figure}

In $^{50}$Ti,
the $4_2^+$ and $3_1^+$ states appear around $3.5$MeV
in our calculation,
whereas they are not seen in experiments.
The occurrence of such additional levels resembles
the case of $^{49}$Sc.
The calculated $0_2^+$ state corresponds well in energy
to the unestablished $0^+$ state.

In $^{54}$Fe,
the number of observed states is larger
than that of the calculated ones,
probably due to the model-space restriction.
We note that the observed $0_2^+$ and $2_2^+$ states,
which are missing in the present calculation,
might belong to a deformed rotational band,
since the energy interval between the two states is much smaller
than that between $0_1^+$ and $2_1^+$.

Within the $k=0$ configuration space,
we have only four states with $J^P=0^+, 2^+, 4^+, 6^+$
in $^{50}$Ti and $^{54}$Fe.
These states consist in the identical energy spectra
between $^{54}$Fe and $^{50}$Ti,
as a consequence of the particle-hole symmetry.
It is noted here that, within the $k=0$ configuration,
$Ex(2_1^+)=1.02$MeV is obtained by the present hamiltonian
for $^{50}$Ti and $^{54}$Fe.
The experimental energies are 1.55MeV for $^{50}$Ti and
1.40MeV for $^{54}$Fe, respectively.
The present calculation within the $k\leq 2$ space
improves the $2_1^+$ levels to a great extent,
somehow overshooting.
Namely, due to the additional pairing correlations
associated with the $k=1$ and $2$ configurations,
the $2_1^+$ energies are raised relative to the ground-state energies.

There are two $4^+$ states around 2.5MeV in $^{52}$Cr.
This approximate degeneracy is reproduced in the present calculation,
where both of the two states predominantly
consist of the $k=0$ configuration.
The correspondence of these states between calculation and experiment
is tentatively made according to the electromagnetic properties,
as will be stated in Sections~\ref{sec:E2} and \ref{sec:M1}.
The observed $0_2^+$ and $2_2^+$ states might be deformed states,
as in the case of $^{54}$Fe.

\subsection{$Z={\em odd},~N=28$ nuclei}
\label{subsec:Zodd-N28}

As shown in Figs.\ref{fig:enV51} and \ref{fig:enMn53},
the present calculation reproduces the observed levels
in $Z={\em odd}$, $N=28$ nuclei
with the energy range of $Ex\lsim 3$MeV,
which is smaller than in the $Z={\em even}$ case.
This is a reasonable consequence,
because ground-state energies in even-even nuclei
are systematically lower than those in neighboring odd nuclei,
owing to the pairing correlation.
Indeed, in the present case,
this is reflected in the higher level densities in odd nuclei
around the ground states.

\begin{figure}
\centering
%\vspace{80mm}
\vspace{30mm}
\caption{\label{fig:enV51}}
\begin{small}
Energy levels of $^{51}$V.
The data are taken from Ref.\cite{ref:NDS51}.
\end{small}
\end{figure}

\begin{figure}
\centering
%\vspace{80mm}
\vspace{30mm}
\caption{\label{fig:enMn53}}
\begin{small}
Energy levels of $^{53}$Mn.
The data are taken from Ref.\cite{ref:NDS53}.
\end{small}
\end{figure}

By comparing Figs.\ref{fig:enV51} and \ref{fig:enMn53},
a notable similarity is found for the lowest five levels
between $^{51}$V and $^{53}$Mn,
analogously to the similarity between $^{50}$Ti and $^{54}$Fe.
Within the space consisting only of the $k=0$ configuration,
this can be understood as a result of the particle-hole symmetry.
The results in Figs.\ref{fig:enV51} and \ref{fig:enMn53}
include the $k=1$ and $2$ configurations,
which break the simple picture of the particle-hole symmetry.
Nevertheless, for the lowest five levels,
the properties of this particle-hole symmetry remain
to a certain extent in the present calculation.
Such properties are seen experimentally also.
On the other hand, the similarity disappears
in higher states with $Ex>2$MeV.
It is noticed that our calculation reproduces several levels
higher than 2MeV.

\subsection{$Z={\em even},~N=29$ nuclei}
\label{subsec:Zeven-N29}

The energy levels of $^{51}$Ti, $^{53}$Cr and $^{55}$Fe are shown
in Figs.\ref{fig:enTi51}, \ref{fig:enCr53} and \ref{fig:enFe55},
respectively.
The good agreement between the calculated and observed ones
is confirmed for $Ex\lsim 3$MeV,
as in the case of the $Z={\em odd}$ nuclei
discussed in Subsection~\ref{subsec:Zodd-N28}.

\begin{figure}
\centering
%\vspace{80mm}
\vspace{30mm}
\caption{\label{fig:enTi51}}
\begin{small}
Energy levels of $^{51}$Ti.
The data are taken from Ref.\cite{ref:NDS51}.
\end{small}
\end{figure}

\begin{figure}
\centering
%\vspace{80mm}
\vspace{30mm}
\caption{\label{fig:enCr53}}
\begin{small}
Energy levels of $^{53}$Cr.
The data are taken from Ref.\cite{ref:NDS53}.
\end{small}
\end{figure}

\begin{figure}
\centering
%\vspace{80mm}
\vspace{30mm}
\caption{\label{fig:enFe55}}
\begin{small}
Energy levels of $^{55}$Fe.
The data are taken from Ref.\cite{ref:NDS55}.
\end{small}
\end{figure}

In $^{51}$Ti there is an unobserved ${7\over 2}^-$ level
at $2.9$MeV (labeled as $[{7\over 2}]_2^-$ in Fig.\ref{fig:enTi51})
in the present calculation.
Concerning this level,
we find a similar pattern in the energy spectra
of $^{53}$Cr and $^{55}$Fe,
in which the $[{7\over 2}]_2^-$ is observed
just above the $[{7\over 2}]_1^-$.
When the good agreement around this level
and the systematics over these three nuclei are considered,
the occurrence of the $[{7\over 2}]_2^-$ state is very likely.
It is of interest whether this level is observed
in the future experiment or not.

We now compare the present results
with those by Horie and Ogawa\cite{ref:Ogawa},
which have been the most successful shell model calculation
in this region.
It should be recalled that only the $k=0$ configuration is included
in the Horie-Ogawa calculation.
Low-lying levels in $N=29$ nuclei are fitted
in determining the proton-neutron interaction
in the Horie-Ogawa calculation.
Up to $Ex=2$MeV, the two results are very similar
except for the $[{7\over 2}]_2^-$ state,
in any of $^{51}$Ti, $^{53}$Cr and $^{55}$Fe
(See Figs.3--5 of Ref.\cite{ref:Ogawa}).
The Horie-Ogawa calculation does not produce
$[{7\over 2}]_2^-$ levels in $^{53}$Cr and $^{55}$Fe,
which have been observed close to $[{7\over 2}]_1^-$ in energy.
This is consistent with our observation that the $k=1$ configuration
dominates the $[{7\over 2}]_2^-$ state of $^{53}$Cr and $^{55}$Fe.
The correspondence between the calculated
and experimental ${7\over 2}^-$ levels
which are nearly degenerate
is made in Fig.\ref{fig:enCr53} and \ref{fig:enFe55}
on the basis of electromagnetic properties.

\subsection{$Z={\em odd},~N=29$ nuclei}
\label{subsec:Zodd-N29}

In Figs.\ref{fig:enSc50}--\ref{fig:enCo56},
we see a good agreement between the calculated energy levels
and experiments
for the $Z={\em odd}$, $N=29$ nuclei,
as far as the energy range of $Ex\lsim 2$MeV is concerned.
This smaller energy range of agreement than that
for even-even or odd nuclei
seems to be associated with the pairing correlation,
as has been stated in Subsection~\ref{subsec:Zodd-N28}.

\begin{figure}
\centering
%\vspace{70mm}
\vspace{30mm}
\caption{\label{fig:enSc50}}
\begin{small}
Energy levels of $^{50}$Sc.
The data are taken from Ref.\cite{ref:NDS50}.
\end{small}
\end{figure}

\begin{figure}
\centering
%\vspace{70mm}
\vspace{30mm}
\caption{\label{fig:enV52}}
\begin{small}
Energy levels of $^{52}$V.
The data are taken from Ref.\cite{ref:NDS52}.
\end{small}
\end{figure}

\begin{figure}
\centering
%\vspace{70mm}
\vspace{30mm}
\caption{\label{fig:enMn54}}
\begin{small}
Energy levels of $^{54}$Mn.
The data are taken from Ref.\cite{ref:NDS54}.
\end{small}
\end{figure}

\begin{figure}
\centering
%\vspace{70mm}
\vspace{30mm}
\caption{\label{fig:enCo56}}
\begin{small}
Energy levels of $^{56}$Co.
The data are taken from Ref.\cite{ref:NDS56}.
\end{small}
\end{figure}

Suppose that we restrict the proton and neutron valence orbits
only to ${\rm 0f}_{7/2}$ and ${\rm 1p}_{3/2}$ respectively,
only four states are possible in $^{50}$Sc;
$J^P=2^+$, $3^+$, $4^+$ and $5^+$.
Indeed, the lowest four states consist predominantly
of that configuration
in the present calculation.
The observed 0.257MeV level is expected to have $J^P=2^+$.
There are several states without the corresponding observed levels
in $^{50}$Sc.
This problem could be accounted similarly as $^{49}$Sc is explained.

Among all nuclei that we discuss in this paper,
the $^{52}$V nucleus is the only one
for which the ground-state spin is not reproduced.
This may not be a serious drawback,
since the lowest three states are very close in energy
and such approximate degeneracy itself is reproduced very well.
Based on the present calculation,
the observed 0.017MeV state appears most likely to be a $2^+$ state.
The tentative assignment of $5^+$ to the 0.023MeV state
seems plausible.

In the Horie-Ogawa results (See Fig.7 of Ref.\cite{ref:Ogawa}),
the $4_1^+$ level has energy very close
to the lowest three levels ($5_1^+$, $3_1^+$ and $2_1^+$),
in contrast to experiment.
The present calculation, on the other hand,
produces a $4_1^+$ level nearly degenerate with $1_1^+$
rather than the lowest three levels,
in agreement with the data.
The $5_2^+$ level, which seems to correspond
to the observed state at $Ex=0.881$MeV,
is dominated by the $k=1$ configuration.
The occurrence of this level is one of the advantages
of the present large-scale calculation,
while this $k=1$ dominance explains
why the Horie-Ogawa interaction is not capable
of describing this state.

It is demonstrated that the lowest five states of $^{54}$Mn
are excellently reproduced.
We confirm that the present result is better
than the result of the Horie-Ogawa calculation
even for these lowest states.
There are several $k=1$ dominant states among the displayed levels;
for instance, $6_1^+$, $1_1^+$, $2_2^+$ and $7_1^+$.
The low energy of the $6_1^+$ cannot be described
by the Horie-Ogawa $k=0$ calculation
(See Fig.8 of Ref.\cite{ref:Ogawa}),
which yields $Ex(6_1^+)\simeq 2$MeV.
Despite a certain underestimation ($\sim 0.4$MeV),
the reproduction of the low $Ex(6_1^+)$ is a prominent advantage
of the present approach.

In $^{56}$Co,
the energy of the $5_1^+$ level shows a certain improvement
from the Horie-Ogawa result (See Fig.9 of Ref.\cite{ref:Ogawa}).
There is an unobserved $2^+$ state around $Ex=1.3$MeV
in the Horie-Ogawa calculation.
Such a state is not seen in the present result.
The $1_1^+$ state has the $k=1$ dominance in the present calculation,
consistently with the failure of the description of this state
in the Horie-Ogawa calculation.

We cannot reproduce the observed low energy of $0_1^+$ in $^{56}$Co.
We should remember, however, that there is no $0^+$ state
in the $k=0$ configuration.
Our calculated $0_1^+$ level at $Ex\simeq 2$MeV,
whose appearance is encouraging to us, may come down
when higher configurations are included.

\subsection{$Z={\em even},~N=30$ nuclei}
\label{subsec:Zeven-N30}

\begin{figure}
\centering
%\vspace{100mm}
\vspace{30mm}
\caption{\label{fig:enTi52}}
\begin{small}
Energy levels of $^{52}$Ti.
The data are taken from Ref.\cite{ref:NDS52}.
\end{small}
\end{figure}

\begin{figure}
\centering
%\vspace{100mm}
\vspace{30mm}
\caption{\label{fig:enCr54}}
\begin{small}
Energy levels of $^{54}$Cr.
The data are taken from Ref.\cite{ref:NDS54}.
\end{small}
\end{figure}

\begin{figure}
\centering
%\vspace{100mm}
\vspace{30mm}
\caption{\label{fig:enFe56}}
\begin{small}
Energy levels of $^{56}$Fe.
The data are taken from Ref.\cite{ref:NDS56}.
\end{small}
\end{figure}

\begin{figure}
\centering
%\vspace{100mm}
\vspace{30mm}
\caption{\label{fig:enNi58}}
\begin{small}
Energy levels of $^{58}$Ni.
The data are taken from Ref.\cite{ref:NDS58}.
\end{small}
\end{figure}

As is already reported in Ref.\cite{ref:Nakada}
and is demonstrated in Fig.\ref{fig:enFe56},
the spectrum of $^{56}$Fe seen in experiments
is excellently reproduced for $Ex \lsim 4$MeV,
apart from the $3^-$ state of 3.07MeV,
which is obviously outside the present configuration space.
Discrepancies in the excitation energies are less than 0.2MeV.
The calculated excitation energies of the yrast states
also agree with the experimental ones up to $J^P=8^+$;
4.75MeV (4.70MeV in experiment) for $7_1^+$,
5.32MeV (5.26MeV) for $8_1^+$.

The energy levels of $^{52}$Ti and $^{54}$Cr are displayed
in Figs.\ref{fig:enTi52} and \ref{fig:enCr54}, respectively.
There are a few calculated but unobserved levels in $^{52}$Ti.
It is not clear, as in the case of $^{49}$Sc,
whether those levels emerge
because the effect of the sd-shell configuration
is too small in the present calculation,
or whether they have not been observed.

The comparison with the data of $^{58}$Ni is presented
in Fig.\ref{fig:enNi58}.
The calculated $0_2^+$ state is much lower than the experimental one.
However, the calculated $0_2^+$ is highly dominated
by the $k=2$ configuration.
It is possible that this remarkable difference of the wavefunction
from the ground state makes it difficult to observe the state.
If we consider the calculated $0_3^+$ state
to correspond to the observed second $0^+$,
the agreement becomes satisfactory.
We then predict a $0^+$ state around 2.2MeV.
The $2_3^+$ state is also dominated by the $k=2$ configuration.
The $B({\rm E2}; 2_3^+ \rightarrow 0_2^+)$ value is fairly large
($200[e^2 {\rm fm}^4]$) in the present calculation,
suggesting a quasi-band structure.
Although, in Fig.\ref{fig:enNi58},
a correspondence of the $2_3^+$ state to the observed $2_3^+$ state
is tentatively indicated based on their energies,
this correspondence can be reconsidered.
The calculated $2_3^+$ state might also remain unobserved.
The comparison between calculation and experiment, however,
is not disturbed by this open question so much.

\subsection{$Z={\em odd},~N=30$ nuclei}
\label{subsec:Zodd-N30}

The calculated and observed energy levels are exhibited
in Figs.\ref{fig:enSc51}--\ref{fig:enCo57},
for $Z={\em odd}$, $N=30$ nuclei.
There are plenty of low-lying states
whose spins and/or parities are not assigned, in these nuclei.
We can give a theoretical suggestion.

\begin{figure}
\centering
%\vspace{80mm}
\vspace{30mm}
\caption{\label{fig:enSc51}}
\begin{small}
Energy levels of $^{51}$Sc.
The data are taken from Ref.\cite{ref:NDS51}.
\end{small}
\end{figure}

\begin{figure}
\centering
%\vspace{80mm}
\vspace{30mm}
\caption{\label{fig:enV53}}
\begin{small}
Energy levels of $^{53}$V.
The data are taken from Ref.\cite{ref:NDS53}.
\end{small}
\end{figure}

\begin{figure}
\centering
%\vspace{80mm}
\vspace{30mm}
\caption{\label{fig:enMn55}}
\begin{small}
Energy levels of $^{55}$Mn.
The data are taken from Ref.\cite{ref:NDS55}.
\end{small}
\end{figure}

\begin{figure}
\centering
%\vspace{80mm}
\vspace{30mm}
\caption{\label{fig:enCo57}}
\begin{small}
Energy levels of $^{57}$Co.
The data are taken from Ref.\cite{ref:NDS57}.
\end{small}
\end{figure}

In $^{51}$Sc, the spin-parity of the fourth lowest state
is ${9\over 2}^-$ in the calculation,
while ${3\over 2}\leq J\leq {7\over 2}$ is suggested in the experiment,
based on the $\log ft$ value of the $\beta$-decay
from $^{51}$Ca\cite{ref:Ca51beta}.
It is desired to specify the spin-parity of this state.
According to the present calculation,
the observed state with $Ex=1.394$MeV should have $J^P={5\over 2}^-$,
which is consistent with the experimental indications.
The spin-parity of the 1.715MeV state
is considered to be ${7\over 2}^-$.

In $^{53}$V,
the calculation strongly suggests the $J^P={9\over 2}^-$ assignment
for the $Ex=1.266$MeV state,
${3\over 2}^-$ for the 1.550MeV state
and ${9\over 2}^-$ for the 1.653MeV state.
None of these suggestions contradict the experiments\cite{ref:NDS53}.

In $^{55}$Mn, the data imply an extraordinary situation
that three states are present around $Ex=1.29$MeV
with surprising degeneracy (${\sl \Delta}E = 0.003$MeV).
Only an ${11\over 2}^-$ state appears in the calculation.

The existence of the ${3\over 2}^-$ state at 1.757MeV in $^{57}$Co
is not explained within the present model space.
This state might be an intruder state dominated
by $k>2$ configuration.

As a whole, the energy levels of the $Z={\em odd}$, $N=30$ nuclei
are pertinently described by the present calculation,
up to $Ex\simeq 2.5$MeV.

\subsection{Summary of energy levels}
\label{subsec:sum-energy}

We summarize this section.
The energy levels of the $20<Z\leq 28$, $28\leq N\leq 30$ nuclei
are well reproduced by the present work,
where the large-scale shell model calculation in the $k\leq 2$ space
is carried out by adopting the Kuo-Brown interaction.
The upper bounds of the agreement with the experiments
are $Ex\simeq 4$MeV for even-even, 2.5MeV for odd-mass
and 2MeV for odd-odd nuclei.
The discrepancy is $\delta E\lsim 0.3$MeV.

Most of the low-lying states are, as are expected,
dominated by the $k=0$ configuration.
In odd-odd nuclei, a certain improvement from the Horie-Ogawa result,
which contains only the $k=0$ configuration,
is achieved even for the lowest-lying $k=0$ dominant states.
Some low-lying states are found to have larger probabilities
of the $k=1$ configuration
than of the $k=0$ configuration;
$3_1^+$ and $4_3^+$ of $^{52}$Cr,
$[{13\over 2}]_1^-$, $[{11\over 2}]_2^-$
and $[{1\over 2}]_1^-$ of $^{53}$Mn,
one of the nearly degenerate ${7\over 2}^-$ of $^{51}$Ti,
$^{53}$Cr and $^{55}$Fe,
$5_2^+$, $3_3^+$ and $4_3^+$ of $^{52}$V,
$6_1^+$, $1_1^+$, $2_2^+$ and $7_1^+$ of $^{54}$Mn,
$1_1^+$ of $^{56}$Co, $4_2^+$ and $2_4^+$ of $^{52}$Ti,
$2_4^+$ and $3_1^+$ of $^{56}$Fe, $4_2^+$ of $^{58}$Ni,
$[{3\over 2}]_1^-$, $[{1\over 2}]_1^-$, $[{7\over 2}]_2^-$
and $[{5\over 2}]_1^-$ of $^{57}$Co.
These levels show good agreement to experiments.
We can expect that the wavefunctions
of the above $k=1$ dominant states
are as good as those of other low-lying states,
since the $k=2$ probabilities of such states are similar
to those of the lowest-lying $k=0$ dominant states.
It is remarked that these $k=1$ dominant states have been
beyond the description by the Horie-Ogawa calculation.

By comparing several effective interactions
derived from the
$G$-matrices\cite{ref:KBpf,ref:Oberlechner,ref:Kuo-pv},
it is inferred that the $G$-matrix interaction
seems to involve an uncertainty of a few hundred keV.
Since this error in the hamiltonian is transferred
to the results of the diagonalization,
the present discrepancy, $\delta E\lsim 0.3$MeV,
appears to be quite reasonable.
Further investigations will be required in order to discern
whether the success of the present calculation is accidental or not.
A test of the wavefunctions through the observation of
electromagnetic properties
is particularly important.

The occurrence of the $0^+$ level
around $Ex\simeq 2.2$MeV in $^{58}$Ni
appears to be a challenging problem.
It is strongly desired to search
for such a $0^+$ level experimentally.
A search for a $0^+$ level around $Ex\simeq 2.3$MeV in $^{56}$Ni,
which emerges in the $k\leq 4$ calculation, is also desired.

\section{E2 properties}
\label{sec:E2}

By using the shell model wavefunctions
we can investigate the electromagnetic properties
of the nuclear states.
In this and the next sections we discuss
the E2 and M1 properties of the $N=28\sim 30$ nuclei.
The purpose of this study is,
as well as to test the shell model wavefunctions,
to see to what extent we understand
the electromagnetic properties of the nuclei in this region
from the nucleonic degrees of freedom
without introducing adjustable parameters.
We employ single-particle parameter sets of the transitions
derived from microscopic theories,
which seem to be the most plausible ones presently available.
We do not attempt to improve those theories
nor to adjust the parameters in this work.
The experimental data are taken
from Refs.\cite{ref:NDS49}--\cite{ref:NDS58}.

\subsection{Single-particle parameters}
\label{subsec:E2sp}

The E2 transition (or moment) is described
by the following one-body operator of nucleons,
\be T({\rm E2}) = \sum_i e_i [r_i^2 Y^{(2)} (\hat{\bm{r}}_i)] ,
   \label{eq:E2gen} \ee
where $i$ is the index of nucleons constituting the nucleus.
Within the framework of the present shell model,
we are dealing only with the valence particles in the pf-shell.
Hence the summation in Eq.(\ref{eq:E2gen}) will be restricted
to the nucleons outside the $^{40}$Ca core,
and the effective charge should be employed
in order to incorporate the core polarization effect.

In many cases the effective charges are fitted phenomenologically
to experimental data
under the assumption of the constancy for many transitions
in one or several nuclei.
In this paper, however,
we derive the effective charges from a microscopic standpoint.
In order to take into account the core polarization effect,
we use the method of Sagawa-Brown\cite{ref:Sagawa-Brown},
which was shown to work well at least for collective transitions
between lowest-lying states in the vicinity of doubly magic core.
Then $e_i$ in Eq.(\ref{eq:E2gen}) is replaced
by $e_\rho^{\rm eff} (j,j')$,
where $\rho$ is the subscript for distinguishing
between protons and neutrons,
and $j$ ($j'$) denotes the initial (final) single-particle orbit
of the $i$-th nucleon.

Although the Kuo-Brown interaction is calculated
by assuming the harmonic oscillator single-particle wavefunctions,
here we adopt the radial part of the single-particle wavefunctions
given by the Hartree-Fock (HF) calculation.
Although we abandon the consistency between them,
this does not give rise to notable differences as shown later.
The HF calculation is carried out
with the SG\Roman{SG2} Skyrme interaction\cite{ref:SG2}
for the $^{56}$Ni core,
assuming the full occupancy of the ${\rm 0f}_{7/2}$ orbit.
The isoscalar (IS) and isovector (IV) quadrupole response functions
are calculated
by the random phase approximation (RPA)\cite{ref:Bertsch-Tsai}.
We searched giant quadrupole resonances (GQR)
in the range of $10<Ex<70$MeV,
and find a single isolated IS-GQR peak at $Ex \simeq 17$MeV
and twenty-four IV-GQR peaks distributed over $Ex = 20 \sim 35$MeV.
The response function is defined by
\be
 S(E) = \sum_n | \bra \omega^{(\lambda)}_n | T^{(\lambda)}
  |0\ket |^2 \delta (E-E_n),
\ee
where the state $|0\ket$ is the HF state,
$|\omega^{(\lambda)}_n\ket$ an excited RPA state
with the angular momentum $\lambda$.
In this case $\lambda=2$,
and $T^{(\lambda)}$ stands for the operator $r^2 Y^{(2)}$
or $r^2 Y^{(2)} \tau_z$.
We show these response functions in Fig.\ref{fig:GQR},
taking into account the escaping width.
The spreading width is not included in this calculation.
By classifying the transition densities at the peak energies
through their shapes,
we select nine most prominent IV-GQR peaks as well as the IS-GQR peak.
The transition strengths are assumed
to concentrate in the selected GQR peaks.
The adopted peaks and the corresponding transition strengths
are listed in Table~\ref{tab:GQR}.

\begin{figure}
\centering
%\vspace{60mm}
\vspace{30mm}
\caption{\label{fig:GQR}}
\begin{small}
Isoscalar and isovector quadrupole response function
from the $^{56}$Ni closed core,
obtained from the HF+RPA calculation
with the SG\Roman{SG2} Skyrme interaction.
\end{small}
\end{figure}

\begin{table}
\centering
\caption{\label{tab:GQR}}
\begin{small}
Excitation energies (MeV) and transition strengths
($e^2 {\rm fm}^4$) of the GQR states
adopted for the calculation of the core polarization effect.
The transition strengths are summed over the states
which possess similar shapes of transition densities.\\
\begin{tabular}{|c|r|r|}
   \hline
     mode &~~~~$Ex$~~~~&~~~~~$B({\rm E2})$~~~~~\\
   \hline
     IS & $16.69$~~ & $800.9~~~$ \\
   \hline
     IV & $21.83~~$ & $29.3~~~$ \\
     & $24.64~~$ & $93.4~~~$ \\
     & $25.60~~$ & $38.7~~~$ \\
     & $27.19~~$ & $9.4~~~$ \\
     & $27.69~~$ & $13.2~~~$ \\
     & $29.77~~$ & $35.2~~~$ \\
     & $30.21~~$ & $31.3~~~$ \\
     & $30.93~~$ & $67.5~~~$ \\
     & $31.86~~$ & $284.3~~~$ \\
   \hline
\end{tabular}
\end{small}
\end{table}

The core polarization effect caused by these GQR states
is incorporated into the single-particle wavefunctions
with the mixing amplitudes evaluated by the perturbation.
When a single-particle state is denoted by $|j\ket$,
the renormalized single-particle state is written as
\be |\tilde j\ket = |j\ket + \sum_{n,j'} a_{n,j'}(j)
 |\omega^{(\lambda)}_n \times j'; j\ket . \ee
Here $a_{n,j'}(j)$ represents the mixing amplitude evaluated
by the perturbation,
\be a_{n,j'}(j) = {{\bra \omega^{(\lambda)}_n \times j'; j|
 V_{\rm res}|j \ket}
\over{\epsilon_j - (\epsilon_{j'}+\omega^{(\lambda)}_n)}},
 \label{eqB:c} \ee
where $V_{\rm res}$ is the residual particle-hole interaction.
Then the renormalized transition density is calculated
by the perturbation as
\bea
\bra \tilde j'\|T^{(\lambda)}(r)\|\tilde j\ket &\simeq&
 \bra j'\|T^{(\lambda)}(r)\|j\ket +
 \sum_n \begin{array}[t]{l} \left\{a_{n,j'}(j) \bra j'\|
  T^{(\lambda)}(r) \|\omega^{(\lambda)}_n \times j'; j\ket \right. \\
 \left. + a_{n,j}(j') \bra \omega^{(\lambda)}_n \times j; j'\|
  T^{(\lambda)}(r) \|j\ket \right\} \end{array} \nonumber \\
&=& \bra j'\|T^{(\lambda)}(r)\|j\ket
 + \sum_n \left\{a_{n,j'}(j) \sqrt{{{2j+1}\over{2\lambda +1}}}
  \bra 0\| T^{(\lambda)}(r) \|\omega^{(\lambda)}_n \ket \right.
   \nonumber \\
&&~~~~~~~~~~~~ \left. + (-)^{\lambda+j-j'} a_{n,j}(j')
 \sqrt{{{2j'+1}\over{2\lambda +1}}} \bra \omega^{(\lambda)}_n \|
  T^{(\lambda)}(r) \|0\ket \right\}.
\label{eqB:TD}
\eea
In the RPA calculation and the evaluation of the mixing amplitudes,
the residual interaction $V_{\rm res}$ is derived
from the SG\Roman{SG2} Skyrme interaction,
consistently with the HF calculation.
After taking expectation values of momentum-dependent terms
with respect to the HF state,
the Skyrme interaction $H_{\rm Sk}$ only depends
on the isoscalar nucleon density $\rho_{\rm IS}(\bm{r})$
and the isovector nucleon density $\rho_{\rm IV}(\bm{r})$.
Therefore the residual interaction is written as
\bea
V_{\rm res}&=&{1\over 2} \left.
 {{\delta^2 H_{\rm Sk}}\over{\delta \rho_{\rm IS}^2}}\right|_0~
 \left[\sum_i \delta (\bm{r}-\bm{r}_i)\right]^2
 + {1\over 2} \left.
  {{\delta^2 H_{\rm Sk}}\over{\delta \rho_{\rm IV}^2}}\right|_0~
 \left[\sum_i \delta (\bm{r}-\bm{r}_i) \tau_z (i)\right]^2 ,
  \label{eqB:Vres}
\eea
where $\mid_0$ means the evaluation by the HF state.
All steps of this renormalization are carried out
at the $^{56}$Ni core.
It should be stressed that this calculation includes
no additional adjustable parameters
besides those contained in the Skyrme interaction.

The reason why we start from the $^{56}$Ni core
is that the nuclei under consideration have masses
closer to $^{56}$Ni than to $^{40}$Ca, as a whole.
The difference in mass number affects $\bra r^2 \ket_{\rm s.p.}$
(expectation values of $r^2$ for the single-particle orbits).
It is noted that, from the $^{56}$Ni core,
we have quadrupole transition strength in the $0\hbar \omega$ space.
This $0\hbar \omega$ strength should be handled principally
within the shell model,
since we have included the $k>0$ configuration.
The quadrupole transition strength within the $0\hbar\omega$ space,
however, seems to be exhausted in $Ex<10$MeV,
since $\hbar\omega \sim 10$MeV at $A=56$.
Indeed we find only a single IS-GQR peak at $Ex\simeq 17$MeV,
which should be dominated by the $2\hbar\omega$ configuration.
We search the RPA peaks for $Ex>10$MeV,
omitting the strength within the $0\hbar\omega$ space.
By this procedure, the E2 transition strength contributed
by the $k>2$ configurations,
which stays within the $0\hbar\omega$ space but is neglected
in the shell model calculation,
cannot be included in the following calculation.
It should be remembered that the E2 properties,
particularly the collective ones,
will provide us with a suitable test for the convergence
with respect to the $k\leq 2$ truncation.

This renormalization procedure naturally introduces
a $j$-dependence into the effective charges.
These $j$-dependent effective charges are shown
in Table~\ref{tab:GQRmix}.
It is seen that the $j$-dependent effective charges are
actually insensitive to $j$.
This fact justifies the usual $j$-independent effective charges
for the E2 transition.
Recall that the $j$-independent charges were adopted
in Ref.\cite{ref:Nakada},
by adjusting the E2 transition probabilities in $^{56}$Fe;
$e^{\rm eff}_\pi = 1.4 e$ and $e^{\rm eff}_\nu = 0.9 e$.
Comparing the microscopic values in Table~\ref{tab:GQRmix}
with those $j$-independent effective charges,
we have a good agreement for proton charges
and a slight suppression for neutron charges.
This implies that the core polarization effect
is slightly underestimated
in both IS and IV modes.
The influence of the $k>2$ configurations within the pf-shell
might be present in $^{56}$Fe
and this influence may account for the difference
in the neutron effective charge.
Note that the single-particle matrix elements of $r^2$
obtained by the HF calculation
is very similar to those in the harmonic oscillator approximation,
when we use the usual oscillator length
$b=A^{1/6}[{\rm fm}]$ at $A=56$.
In practice, the ratios of the HF matrix elements
to the harmonic oscillator ones are 0.93$\sim$1.05.
Although there ought to be a certain decrease
of the size of the single-particle wavefunctions
as $A$ becomes smaller,
we will neglect this effect in the following calculation.
This decrease amounts only to a 7\% reduction of $B({\rm E2})$
at $A=50$,
by estimating it under the assumption of
$\bra r^2 \ket_{\rm s.p.} \propto b^2 \propto A^{1/3}$.

\begin{table}
\centering
\caption{\label{tab:GQRmix}}
\begin{small}
E2 effective charges evaluated by the method of Sagawa-Brown.\\
\begin{tabular}{|c|c||r|r|}
   \hline
     $j$ & $j'$ & $e^{\rm eff}_\pi (j,j')$
      & $e^{\rm eff}_\nu (j,j')$ \\
   \hline
     ${\rm 0f}_{7/2}$ & ${\rm 0f}_{7/2}$ & $1.354~~$ & $0.677~~$ \\
     ${\rm 0f}_{7/2}$ & ${\rm 0f}_{5/2}$ & $1.482~~$ & $0.841~~$ \\
     ${\rm 0f}_{7/2}$ & ${\rm 1p}_{3/2}$ & $1.488~~$ & $0.788~~$ \\
     ${\rm 0f}_{5/2}$ & ${\rm 0f}_{5/2}$ & $1.336~~$ & $0.643~~$ \\
     ${\rm 0f}_{5/2}$ & ${\rm 1p}_{3/2}$ & $1.436~~$ & $0.712~~$ \\
     ${\rm 0f}_{5/2}$ & ${\rm 1p}_{1/2}$ & $1.411~~$ & $0.690~~$ \\
     ${\rm 1p}_{3/2}$ & ${\rm 1p}_{3/2}$ & $1.341~~$ & $0.601~~$ \\
     ${\rm 1p}_{3/2}$ & ${\rm 1p}_{1/2}$ & $1.343~~$ & $0.604~~$ \\
   \hline
\end{tabular}
\end{small}
\end{table}

\clearpage
\subsection{$Z={\em even},~N=28$ nuclei}
\label{subsec:E2_Zeven-N28}

In Table~\ref{tab:E2_ZeN28}, the calculated E2 properties
are compared with the measured ones,
for $Z={\em even}$, $N=28$ nuclei.

We first see $B({\rm E2}; 2_1^+ \rightarrow 0_1^+)$ in these nuclei.
In the description only with the $k=0$ configuration,
the identical transition rates are predicted
between $^{50}$Ti and $^{54}$Fe
due to the particle-hole conjugation,
as far as we neglect the change of the size
of the ${\rm 0f}_{7/2}$ single-particle wavefunction.
The present $k\leq 2$ calculation inherits this feature
to a certain extent.
Provided that the seniority is conserved as discussed
in the work of Horie-Ogawa\cite{ref:Ogawa},
we also have the same $B({\rm E2})$ value for $^{52}$Cr
as for $^{50}$Ti or $^{54}$Fe,
within the $k=0$ space.
In the present calculation,
$B({\rm E2}; 2_1^+ \rightarrow 0_1^+)$ is distinctly enhanced
in $^{52}$Cr,
due to seniority mixing\cite{ref:Vervier}
and the influence of the $k>0$ configurations.

On the other hand,
the $B({\rm E2})$ values of $^{54}$Fe is
about twice as large as that of $^{50}$Ti.
The $B({\rm E2})$ value in $^{52}$Cr is enhanced
from that in $^{50}$Ti,
while it is close to the value in $^{54}$Fe.
Although the present calculation is far from
reproducing the data for $^{50}$Ti and $^{54}$Fe,
which has been a problem for a few decades\cite{ref:Vervier},
the result in $^{52}$Cr yields a good agreement.
As has been discussed in Subsection~\ref{subsec:config},
the $k\leq 2$ truncation might be insufficient
in reproducing some properties of $^{54}$Fe.
The $k>2$ influence may account for the underestimate.
In $^{50}$Ti, the sd-shell contribution might be considerable.
It is not easy to say whether the stronger mixing
of the sd-shell configurations
leads to an overestimate or an underestimate,
because the adopted effective charges contain
the excitation from the sd-shell.
The success in $^{52}$Cr, however, may indicate
that the amount of core polarization is appropriately evaluated.
Since effects of the $k>2$ configurations are not included
in calculating the effective charges,
the wavefunctions look almost convergent
by the $k\leq 2$ truncation in this nucleus.
It would be impossible to reproduce
the rapid change of the $B({\rm E2})$ values
without either more precise wavefunctions
including the $k>2$ configurations
and the sd-shell excitations explicitly,
or a strong particle number dependence of effective charges.

In $^{52}$Cr, there are two $4^+$ states around 2.5MeV.
Judging from $B({\rm E2}; 4^+ \rightarrow 2_1^+)$
and $B({\rm E2}; 6_1^+ \rightarrow 4^+)$,
the lower $4^+$ level in the calculation corresponds
rather to the observed $4_2^+$ state,
whereas the higher one to the observed $4_1^+$ state.
This inversion has been taken into account
in Fig.\ref{fig:enCr52} and Table~\ref{tab:E2_ZeN28}.
Through this assignment,
we recognize a good agreement between
the calculated and measured $B({\rm E2})$ values.

In the $k=0$ space, the calculated quadrupole moments
of the $2_1^+$ states
should have the same magnitudes in $^{50}$Ti and $^{54}$Fe,
while the signs are opposite to each other:
positive in $^{50}$Ti and negative in $^{54}$Fe.
Although the signs are maintained,
the magnitudes are different in the $k\leq 2$ calculation.
The calculated quadrupole moment in $^{54}$Fe is almost
twice larger than that in $^{50}$Ti.
The experimental data contain too large errors
to confirm this tendency.
In $^{52}$Cr, $Q(2_1^+)$ should be zero
under the seniority conservation in the $k=0$ space.
This obviously contradicts with the experiment,
which gives a definitely negative value.
The present calculation reproduces this property.

The $B({\rm E2}; 2_1^+ \rightarrow 0_1^+)$ value of $^{56}$Ni
is already shown in Table~\ref{tab:Ni56}.
The result of the $k\leq 4$ calculation,
as well as that of the $k\leq 2$ calculation,
are compared with the experimental value
exhibited in Ref.\cite{ref:NDS56}.
Both the $k\leq 2$ and $k\leq 4$ calculations give similar values,
in good agreement with the datum.

\begin{table*}
\centering
\caption{\label{tab:E2_ZeN28}}
\begin{small}
$B({\rm E2})$ values ($e^2 {\rm fm}^4$) or E2 static moments
($e {\rm fm}^2$) in $Z={\em even}$, $N=28$ nuclei.
The expression $i$ ($f$) for the second (third) column
denotes initial (final) state.
The `Cal.' values are obtained by the present calculation.
The experimental data (Exp.) are taken
from Refs.\cite{ref:NDS50,ref:NDS52,ref:NDS54}.\\
\begin{tabular}{|c||c|c||r@{.}l|r@{.}l@{$\pm$}r@{.}l|}
   \hline
     nucl. & $i$ & $f$ & \multicolumn{2}{c|}{Cal.}
      &  \multicolumn{4}{c|}{Exp.} \\
   \hline
     $^{50}$Ti & $2_1^+$ & $0_1^+$ & $88$&$0~$ & $58$&&$~~~9$& \\
     & $4_1^+$ & $2_1^+$ & $86$&$8~$ & $60$&&$13$& \\
     & $6_1^+$ & $4_1^+$ & $40$&$1~$ & $34$&$4$&$1$&$4$ \\
     & $0_2^+$ & $2_1^+$ & $1$&$6~$ & $18$&&$9$& \\
     & $2_1^+$ & $2_1^+$ & $9$&$0^{*}$
      & $8$&&$16$&\multicolumn{1}{l|}{$^{*}$} \\
   \hline
     $^{52}$Cr & $2_1^+$ & $0_1^+$ & $102$&$7~$ & $131$&&$6$& \\
     & $4_1^+$ & $2_1^+$ & $87$&$9~$ & $84$&&$22$& \\
     & $4_2^+$ & $2_1^+$ & $19$&$6~$ & $53$&&$14$& \\
     & $2_2^+$ & $0_1^+$ & $0$&$05$ & $0$&$06$&$0$&$05$ \\
     & $2_2^+$ & $2_1^+$ & $113$&$4~$ & $150$&&$35$& \\
   \hline
\end{tabular}
\begin{tabular}{|c||c|c||r@{.}l|r@{.}l@{$\pm$}r@{.}l|}
   \hline
     nucl. & $i$ & $f$ & \multicolumn{2}{c|}{Cal.}
      &  \multicolumn{4}{c|}{Exp.} \\
   \hline
     $^{52}$Cr & $6_1^+$ & $4_1^+$ & $74$&$4~$ & $59$&&$2$& \\
     & $6_1^+$ & $4_2^+$ & $13$&$2~$ & $29$&$7$&$1$&$2$ \\
     & $4_3^+$ & $4_2^+$ & $0$&$5~$ & $127$&&$8$& \\
     & $3_1^+$ & $4_2^+$ & $0$&$2~$ & $7$&&$5$& \\
     & $2_1^+$ & $2_1^+$ & $- 5$&$8^{*}$
      & $-14$&&$8$&\multicolumn{1}{l|}{$^{*}$} \\
   \hline
     $^{54}$Fe & $2_1^+$ & $0_1^+$ & $73$&$3~$ & $128$&$5$&$4$&$8$ \\
     & $4_1^+$ & $2_1^+$ & $68$&$6~$ & \multicolumn{4}{c|}{---} \\
     & $2_1^+$ & $2_1^+$ & $-16$&$4^{*}$
      & $- 5$&&$14$&\multicolumn{1}{l|}{$^{*}$} \\
   \hline
\end{tabular}
\\
$^*$) Quadrupole moment.
\end{small}
\end{table*}

\clearpage
\subsection{$Z={\em odd},~N=28$ nuclei}
\label{subsec:E2_Zodd-N28}

Table~\ref{tab:E2_ZoN28} shows the E2 quantities
in $Z={\em odd}$, $N=28$ nuclei.

As in the case of $^{50}$Ti and $^{54}$Fe,
$^{51}$V and $^{53}$Mn are closely related
through the particle-hole conjugation
within the $k=0$ space.
If the seniority is a good quantum number,
any corresponding transitions between low-lying states
have the same $B({\rm E2})$ value in these two nuclei,
while the quadrupole moments of the corresponding states
have identical magnitudes but opposite signs.
The present calculation maintains this nature
for many E2 transitions.
The transition from $[{5\over 2}]^-_1$ to $[{7\over 2}]^-_1$
is a typical example,
which is also confirmed by the experiment.
On the other hand, the most significant deviation is seen
in $B({\rm E2}; [{3\over 2}]^-_1 \rightarrow [{5\over 2}]^-_1)$.
The $B({\rm E2})$ value of this transition in $^{51}$V is
almost five times larger
than that in $^{53}$Mn.
These properties are reproduced quite well
by the present $k\leq 2$ calculation.

The results obtained in the present work
are remarkably good in $^{51}$V.
There are certain discrepancies in $^{53}$Mn,
which cannot be removed
without discarding the particle-hole symmetry.
This could be an influence of the $k>2$ configurations.

\begin{table*}
\centering
\caption{\label{tab:E2_ZoN28}}
\begin{small}
$B({\rm E2})$ values ($e^2 {\rm fm}^4$) or E2 static moments
($e {\rm fm}^2$) in $Z={\em odd}$, $N=28$ nuclei.
The experimental data are taken
from Refs.\cite{ref:NDS49,ref:NDS51,ref:NDS53,ref:NDS55}.\\
\begin{tabular}{|c||c|c||r@{.}l|r@{.}l@{$\pm$}r@{.}l|}
   \hline
     nucl. & $i$ & $f$ & \multicolumn{2}{c|}{Cal.}
      &  \multicolumn{4}{c|}{Exp.} \\
   \hline
     $^{49}$Sc & $[{7\over 2}]_1^-$ & $[{7\over 2}]_1^-$
      & $- 20$&$9^{*}$ & \multicolumn{4}{c|}{---} \\
   \hline
     $^{51}$V & $[{5\over 2}]_1^-$ & $[{7\over 2}]_1^-$
      & $188$&$3~$ & $169$&&$~~34$& \\
     & $[{3\over 2}]_1^-$ & $[{7\over 2}]_1^-$
      & $75$&$6~$ & $89$&&$10$& \\
     & $[{3\over 2}]_1^-$ & $[{5\over 2}]_1^-$
      & $101$&$1~$ & $118$&&$16$& \\
     & $[{11\over 2}]_1^-$ & $[{7\over 2}]_1^-$
      & $95$&$2~$ & $125$&&$21$& \\
     & $[{9\over 2}]_1^-$ & $[{7\over 2}]_1^-$
      & $31$&$6~$ & $37$&&$6$& \\
     & $[{9\over 2}]_1^-$ & $[{5\over 2}]_1^-$
      & $29$&$9~$ & $29$&&$5$& \\
     & $[{15\over 2}]_1^-$ & $[{11\over 2}]_1^-$
      & $71$&$6~$ & $66$&&$8$& \\
     & $[{7\over 2}]_1^-$ & $[{7\over 2}]_1^-$
      & $- 5$&$7^{*}$ & $-4$&$3$&$0$&$5^{*}$ \\
   \hline
\end{tabular}
\begin{tabular}{|c||c|c||r@{.}l|r@{.}l@{$\pm$}r@{.}l|}
   \hline
     nucl. & $i$ & $f$ & \multicolumn{2}{c|}{Cal.}
      &  \multicolumn{4}{c|}{Exp.} \\
   \hline
     $^{53}$Mn & $[{5\over 2}]_1^-$ & $[{7\over 2}]_1^-$
      & $180$&$5~$ & $165$&&$35$& \\
     & $[{3\over 2}]_1^-$ & $[{7\over 2}]_1^-$
      & $62$&$0~$ & $158$&&$13$& \\
     & $[{3\over 2}]_1^-$ & $[{5\over 2}]_1^-$
      & $22$&$1~$ & $18$&&$6$& \\
     & $[{11\over 2}]_1^-$ & $[{7\over 2}]_1^-$
      & $83$&$4~$ & $151$&&$21$& \\
     & $[{9\over 2}]_1^-$ & $[{7\over 2}]_1^-$
      & $35$&$4~$ & $83$&&$11$& \\
     & $[{9\over 2}]_1^-$ & $[{5\over 2}]_1^-$
      & $33$&$9~$ & $44$&&$7$& \\
     & $[{1\over 2}]_1^-$ & $[{5\over 2}]_1^-$
      & $0$&$1~$ & $180$&&$140$& \\
     & $[{15\over 2}]_1^-$ & $[{11\over 2}]_1^-$
      & $47$&$7~$ & $64$&&$10$& \\
     & $[{7\over 2}]_1^-$ & $[{7\over 2}]_1^-$
      & $7$&$5^{*}$ & \multicolumn{4}{c|}{---} \\
   \hline
     $^{55}$Co & $[{7\over 2}]_1^-$ & $[{7\over 2}]_1^-$
      & $20$&$6^{*}$ & \multicolumn{4}{c|}{---} \\
   \hline
\end{tabular}
\\
$^*$) Quadrupole moment.
\end{small}
\end{table*}

\clearpage
\subsection{$Z={\em even},~N=29$ nuclei}
\label{subsec:E2_Zeven-N29}

We show the E2 properties of $Z={\em even}$, $N=29$ nuclei
in Table~\ref{tab:E2_ZeN29}.

Generally speaking, large errors in experiments
make any precise comparison difficult.
We only discuss the nearly degenerate ${7\over 2}^-$ states.
Let us consider
$B({\rm E2}; [{7\over 2}]^-_1 \rightarrow [{3\over 2}]^-_1)$
and $B({\rm E2}; [{7\over 2}]^-_2 \rightarrow [{3\over 2}]^-_1)$.
The measured
$B({\rm E2}; [{7\over 2}]^-_1 \rightarrow [{3\over 2}]^-_1)$ value
points out the collectivity of this transition.
In $^{53}$Cr, the calculated transition strength
from the higher ${7\over 2}^-$ level
agrees with the measured value much better
than that from the lower ${7\over 2}^-$ level.
We therefore conclude that, in $^{53}$Cr,
the calculated higher ${7\over 2}^-$ level
should correspond to the observed $[{7\over 2}]^-_1$ state,
whereas the lower one to the observed $[{7\over 2}]^-_2$ state.
This assignment has been taken into account in Fig.\ref{fig:enCr53}
and Table~\ref{tab:E2_ZeN29}.
The transition rates from $[{11\over 2}]^-_1$ and $[{9\over 2}]^-_1$
to these ${7\over 2}^-$ states
support this conclusion.
The transition probabilities from the ${7\over 2}^-$ states
to $[{5\over 2}]^-_1$
also prefer this reversed correspondence.
On the other hand, in $^{55}$Fe,
as far as we can judge from the transitions to the ground states,
the energy sequence of the calculated $[{7\over 2}]^-_1$
and $[{7\over 2}]^-_2$ states
seems to be correctly reproduced, as is assigned presently.

\begin{table*}
\centering
\caption{\label{tab:E2_ZeN29}}
\begin{small}
$B({\rm E2})$ values ($e^2 {\rm fm}^4$) or E2 static moments
($e {\rm fm}^2$) in $Z={\em even}$, $N=29$ nuclei.
The experimental data are taken
from Refs.\cite{ref:NDS51,ref:NDS53,ref:NDS55,ref:NDS57}.\\
\begin{tabular}{|c||c|c||r@{.}l|r@{.}l@{$\pm$}r@{.}l|}
   \hline
     nucl. & $i$ & $f$ & \multicolumn{2}{c|}{Cal.}
      &  \multicolumn{4}{c|}{Exp.} \\
   \hline
     $^{51}$Ti & $[{1\over 2}]_1^-$ & $[{3\over 2}]_1^-$
      & $102$&$5~$ & \multicolumn{4}{c|}{---} \\
     & $[{7\over 2}]_1^-$ & $[{3\over 2}]_1^-$
      & $82$&$7~$ & $230$&&$~200$& \\
     & $[{5\over 2}]_1^-$ & $[{3\over 2}]_1^-$
      & $96$&$4~$ & $350$&&$110$& \\
     & $[{5\over 2}]_2^-$ & $[{3\over 2}]_1^-$
      & $1$&$0~$ & $49$&&$10$& \\
     & $[{5\over 2}]_2^-$ & $[{1\over 2}]_1^-$
      & $80$&$6~$ & $100$&&$21$& \\
     & $[{5\over 2}]_2^-$ & $[{7\over 2}]_1^-$
      & $20$&$4~$ & $382$&&$79$& \\
     & $[{3\over 2}]_2^-$ & $[{3\over 2}]_1^-$
      & $37$&$5~$ & $376$&&$12$& \\
     & $[{11\over 2}]_1^-$ & $[{7\over 2}]_1^-$
      & $110$&$1~$ & $95$&&$17$& \\
     & $[{15\over 2}]_1^-$ & $[{11\over 2}]_1^-$
      & $59$&$8~$ & $62$&&$24$& \\
     & $[{3\over 2}]_1^-$ & $[{3\over 2}]_1^-$
      & $- 11$&$3^{*}$ & \multicolumn{4}{c|}{---} \\
   \hline
     $^{53}$Cr & $[{1\over 2}]_1^-$ & $[{3\over 2}]_1^-$
      & $209$&$0~$ & \multicolumn{4}{c|}{---} \\
     & $[{5\over 2}]_1^-$ & $[{3\over 2}]_1^-$
      & $44$&$8~$ & $169$&&$28$& \\
     & $[{7\over 2}]_1^-$ & $[{3\over 2}]_1^-$
      & $146$&$1~$ & $124$&&$11$& \\
     & $[{7\over 2}]_1^-$ & $[{5\over 2}]_1^-$
      & $10$&$3~$ & $76$&&$19$& \\
     & $[{7\over 2}]_2^-$ & $[{3\over 2}]_1^-$
      & $0$&$2~$ & $0$&$26$&$0$&$06$ \\
     & $[{7\over 2}]_2^-$ & $[{5\over 2}]_1^-$
      & $0$&$0003$ & $1$&$9$&$1$&$7$ \\
     & $[{5\over 2}]_2^-$ & $[{3\over 2}]_1^-$
      & $46$&$9~$ & $43$&&$19$& \\
     & $[{5\over 2}]_2^-$ & $[{7\over 2}]_1^-$
      & $27$&$7~$ & $470$&&$470$& \\
     & $[{11\over 2}]_1^-$ & $[{7\over 2}]_1^-$
      & $132$&$9~$ & $110$&&$12$& \\
     & $[{11\over 2}]_1^-$ & $[{7\over 2}]_2^-$
      & $5$&$2~$ & \multicolumn{4}{c|}{---} \\
   \hline
\end{tabular}
\begin{tabular}{|c||c|c||r@{.}l|r@{.}l@{$\pm$}r@{.}l|}
   \hline
     nucl. & $i$ & $f$ & \multicolumn{2}{c|}{Cal.}
      &  \multicolumn{4}{c|}{Exp.} \\
   \hline
     $^{53}$Cr & $[{11\over 2}]_2^-$ & $[{7\over 2}]_1^-$
      & $18$&$2~$ & \multicolumn{4}{c|}{---} \\
     & $[{11\over 2}]_2^-$ & $[{7\over 2}]_2^-$
      & $66$&$1~$ & \multicolumn{4}{c|}{---} \\
     & $[{9\over 2}]_1^-$ & $[{7\over 2}]_1^-$
      & $0$&$8~$ & \multicolumn{4}{c|}{---} \\
     & $[{9\over 2}]_1^-$ & $[{7\over 2}]_2^-$
      & $306$&$8~$ & $270$&&$110$& \\
     & $[{3\over 2}]_2^-$ & $[{3\over 2}]_1^-$
      & $21$&$9~$ & $130$&&$260$& \\
     & $[{15\over 2}]_1^-$ & $[{11\over 2}]_1^-$
      & $90$&$9~$ & $30$&&$12$& \\
     & $[{15\over 2}]_1^-$ & $[{11\over 2}]_2^-$
      & $6$&$1~$ & $17$&&$11$& \\
     & $[{3\over 2}]_1^-$ & $[{3\over 2}]_1^-$
      & $- 15$&$4^{*}$ & $-15$&&$5$&\multicolumn{1}{l|}{$^{*}$} \\
   \hline
     $^{55}$Fe & $[{1\over 2}]_1^-$ & $[{3\over 2}]_1^-$
      & $205$&$7~$ & \multicolumn{4}{c|}{---} \\
     & $[{5\over 2}]_1^-$ & $[{3\over 2}]_1^-$
      & $30$&$3~$ & $11$&$2$&$~~~5$&$0$ \\
     & $[{5\over 2}]_1^-$ & $[{1\over 2}]_1^-$
      & $92$&$4~$ & $37$&&$24$& \\
     & $[{7\over 2}]_1^-$ & $[{3\over 2}]_1^-$
      & $118$&$7~$ & $62$&&$50$& \\
     & $[{7\over 2}]_1^-$ & $[{5\over 2}]_1^-$
      & $8$&$3~$ & $11$&&$12$& \\
     & $[{7\over 2}]_2^-$ & $[{3\over 2}]_1^-$
      & $0$&$6~$ & $1$&$18$&$0$&$06$ \\
     & $[{7\over 2}]_2^-$ & $[{5\over 2}]_1^-$
      & $0$&$005$ & $1$&$5$&$1$&$9$ \\
     & $[{9\over 2}]_1^-$ & $[{7\over 2}]_2^-$
      & $203$&$4~$ & $92$&&$31$& \\
     & $[{9\over 2}]_2^-$ & $[{5\over 2}]_1^-$
      & $88$&$6~$ & $170$&&$150$& \\
     & $[{3\over 2}]_1^-$ & $[{3\over 2}]_1^-$
      & $- 14$&$3^{*}$ & \multicolumn{4}{c|}{---} \\
   \hline
     $^{57}$Ni & $[{5\over 2}]_1^-$ & $[{3\over 2}]_1^-$
      & $9$&$5~$ & $32$&$6$&$7$&$8$ \\
     & $[{1\over 2}]_1^-$ & $[{3\over 2}]_1^-$
      & $48$&$3~$ & \multicolumn{4}{l|}{$~<3900.$} \\
     & $[{3\over 2}]_1^-$ & $[{3\over 2}]_1^-$
      & $- 7$&$6^{*}$ & \multicolumn{4}{c|}{---} \\
   \hline
\end{tabular}
\\
$^*$) Quadrupole moment.
\end{small}
\end{table*}

\clearpage
\subsection{$Z={\em odd},~N=29$ nuclei}
\label{subsec:E2_Zodd-N29}

In Table~\ref{tab:E2_ZoN29},
the E2 properties are shown for $Z={\em odd}$, $N=29$ nuclei.

We can notice the overall agreement
between the present calculation and the data,
though disagreement in the order of magnitude is seen
in $B({\rm E2}; 1_1^+ \rightarrow 2_1^+)$ in $^{54}$Mn,
$B({\rm E2}; 3_2^+ \rightarrow 3_1^+)$,
$B({\rm E2}; 1_1^+ \rightarrow 3_1^+)$
and $B({\rm E2}; 6_1^+ \rightarrow 5_1^+)$ in $^{56}$Co.
As will be discussed later,
the M1 properties of the relevant states are reproduced fairly well.
Therefore the shell model wavefunctions of these states
are probably adequate.
These discrepancies may originate from the influence
of the $k>2$ configurations,
or from the variation of the core polarization effect,
which is possible since the initial states of
the relevant transitions have rather high energy.

\begin{table*}
\centering
\caption{\label{tab:E2_ZoN29}}
\begin{small}
$B({\rm E2})$ values ($e^2 {\rm fm}^4$) or E2 static moments
($e {\rm fm}^2$) in $Z={\em odd}$, $N=29$ nuclei.
The experimental data are taken
from Refs.\cite{ref:NDS50,ref:NDS52,ref:NDS54,ref:NDS56}.\\
\begin{tabular}{|c||c|c||r@{.}l|r@{.}l@{$\pm$}r@{.}l|}
   \hline
     nucl. & $i$ & $f$ & \multicolumn{2}{c|}{Cal.}
      &  \multicolumn{4}{c|}{Exp.} \\
   \hline
     $^{50}$Sc & $3_1^+$ & $5_1^+$ & $24$&$0~$
      & \multicolumn{4}{c|}{---} \\
     & $4_1^+$ & $5_1^+$ & $3$&$5~$
      & \multicolumn{4}{c|}{---} \\
     & $5_1^+$ & $5_1^+$ & $-25$&$8^{*}$
      & \multicolumn{4}{c|}{---} \\
   \hline
     $^{52}$V & $2_1^+$ & $3_1^+$ & $62$&$0~$
      & \multicolumn{4}{l|}{$~~<~9.\times 10^5$} \\
     & $5_1^+$ & $3_1^+$ & $75$&$5~$ & \multicolumn{4}{c|}{---} \\
     & $1_1^+$ & $3_1^+$ & $96$&$3~$ & \multicolumn{4}{c|}{---} \\
     & $4_1^+$ & $3_1^+$ & $31$&$0~$ & \multicolumn{4}{c|}{---} \\
     & $3_1^+$ & $3_1^+$ & $3$&$8^{*}$ & \multicolumn{4}{c|}{---} \\
   \hline
     $^{54}$Mn & $2_1^+$ & $3_1^+$ & $6$&$8~$
      & \multicolumn{4}{c|}{---} \\
     & $4_1^+$ & $3_1^+$ & $83$&$9~$ & $29$&&$~~11$& \\
     & $5_1^+$ & $3_1^+$ & $58$&$4~$
      & $109$&&\multicolumn{2}{c|}{} \\
     & $5_1^+$ & $4_1^+$ & $118$&$2~$
      & $280$&&\multicolumn{2}{l|}{$^{~720.}_{~220.}$} \\
     & $3_2^+$ & $3_1^+$ & $0$&$4~$
      & \multicolumn{4}{l|}{$~~<~280.$} \\
     & $3_2^+$ & $2_1^+$ & $172$&$6~$ & $194$&&$97$& \\
     & $3_2^+$ & $4_1^+$ & $90$&$8~$
      & $120$&&\multicolumn{2}{l|}{$^{~680.}_{~~90.}$} \\
     & $4_2^+$ & $5_1^+$ & $35$&$9~$
      & $50$&&\multicolumn{2}{l|}{$^{~150.}_{~~50.}$} \\
     & $3_3^+$ & $2_1^+$ & $20$&$3~$
      & $17$&&\multicolumn{2}{l|}{$^{~~67.}_{~~12.}$} \\
     & $3_3^+$ & $4_1^+$ & $31$&$6~$
      & $22$&&\multicolumn{2}{l|}{$^{~~39.}_{~~17.}$} \\
   \hline
\end{tabular}
\begin{tabular}{|c||c|c||r@{.}l|r@{.}l@{$\pm$}r@{.}l|}
   \hline
     nucl. & $i$ & $f$ & \multicolumn{2}{c|}{Cal.}
      &  \multicolumn{4}{c|}{Exp.} \\
   \hline
     $^{54}$Mn & $6_1^+$ & $5_1^+$ & $0$&$003$
      & \multicolumn{4}{l|}{$~~<~~~0.02$} \\
     & $1_1^+$ & $2_1^+$ & $2$&$4~$ & $145$&&$97$& \\
     & $3_1^+$ & $3_1^+$ & $31$&$6^{*}$
      & $33$&&$3$&\multicolumn{1}{l|}{$^{*}$} \\
   \hline
     $^{56}$Co & $3_1^+$ & $4_1^+$ & $47$&$1~$
      & \multicolumn{4}{l|}{$~~>~~~3.6$} \\
     & $5_1^+$ & $4_1^+$ & $58$&$3~$
      & $760$&&\multicolumn{2}{l|}{$^{1010.}_{~510.}$} \\
     & $4_2^+$ & $4_1^+$ & $18$&$5~$
      & \multicolumn{4}{l|}{$~~<~~70.$} \\
     & $4_2^+$ & $3_1^+$ & $40$&$2~$
      & \multicolumn{4}{l|}{$~~<~~52.$} \\
     & $2_1^+$ & $4_1^+$ & $7$&$1~$ & $25$&$5$&$8$&$9$ \\
     & $2_1^+$ & $3_1^+$ & $3$&$5~$
      & $9$&&\multicolumn{2}{l|}{$^{~~18.}_{~~~9.}$} \\
     & $5_2^+$ & $4_1^+$ & $8$&$7~$
      & $13$&&\multicolumn{2}{l|}{$^{~~23.}_{~~10.}$} \\
     & $3_2^+$ & $4_1^+$ & $5$&$2~$ & $10$&$2$&$8$&$9$ \\
     & $3_2^+$ & $3_1^+$ & $3$&$6~$ & $89$&&$51$& \\
     & $3_2^+$ & $4_2^+$ & $52$&$2~$
      & $382$&&\multicolumn{2}{l|}{$^{~~64.}_{~~38.}$} \\
     & $1_1^+$ & $3_1^+$ & $0$&$3~$ & $25$&&$13$& \\
     & $7_1^+$ & $5_1^+$ & $0$&$3~$
      & \multicolumn{4}{l|}{$~~<~~32.$} \\
     & $6_1^+$ & $5_1^+$ & $46$&$6~$
      & $0$&$5$&\multicolumn{2}{l|}{$^{~~~1.7}_{~~~0.5}$} \\
     & $4_1^+$ & $4_1^+$ & $22$&$3^{*}$ & \multicolumn{4}{c|}{---} \\
   \hline
\end{tabular}
\\
$^*$) Quadrupole moment.
\end{small}
\end{table*}

\clearpage
\subsection{$Z={\em even},~N=30$ nuclei}
\label{subsec:E2_Zeven-N30}

Table~\ref{tab:E2_ZeN30} exhibits
the $B({\rm E2})$ values and quadrupole moments
in $Z={\em even}$, $N=30$ nuclei.

In, a good agreement is obtained.
The only problem found in $^{52}$Ti and $^{54}$Cr
is the transition from $2_3^+$ to $2_1^+$ in both nuclei.
It should be noted
that for the $\gamma$-transition from $2_3^+$ to $2_1^+$ in $^{54}$Cr,
a different E2/M1 mixing ratio from the previous value
has recently been reported\cite{ref:Collins}.
The $B({\rm E2})$ value should be reduced if this new value is adopted.

In $^{56}$Fe, it is found that the present calculation
tends to underestimate the $B({\rm E2})$ values.
The $k>2$ configurations will lead to expected larger collectivity.
Keeping this in mind, we can recognize overall agreement
also in this nucleus.
The $6_2^+$ state has a large E2 transition probability to $4_1^+$,
indicating a certain collectivity of the state.
This $B({\rm E2})$ value is even larger
than $B({\rm E2}; 6_1^+ \rightarrow 4_1^+)$.
This feature is properly reproduced in the present calculation.
There are certain discrepancies
in $B({\rm E2}; 1_1^+ \rightarrow 2_1^+)$,
$B({\rm E2}; 4_2^+ \rightarrow 4_1^+)$,
$B({\rm E2}; 6_1^+ \rightarrow 4_2^+)$
and $B({\rm E2}; 6_2^+ \rightarrow 6_1^+)$,
whose initial states have rather high energy.

The $B({\rm E2}; 2_1^+ \rightarrow 0_1^+)$ is
quite underestimated in $^{58}$Ni.
It has been known that larger effective charges
than those used in systematics are needed
in order to adjust $B({\rm E2}; 2_1^+ \rightarrow 0_1^+)$
in $^{58}$Ni\cite{ref:Cohen-Auerbach}.
Shimizu and Arima described the $B({\rm E2})$ value
by taking into account the $k=1$ and $2$ configurations,
based on the pseudo-SU(3) picture\cite{ref:pseudoSU3}.
Although the present calculation includes
the $k=1$ and $2$ configurations fully,
anomalously larger effective charges are still necessary
to reproduce the $B({\rm E2})$ value in $^{58}$Ni,
as in the case of $^{54}$Fe.

As mentioned in Section~\ref{sec:energy},
the calculated $0_3^+$ state of $^{58}$Ni
corresponds rather well in energy
to the second lowest $0^+$ state observed so far.
The $B({\rm E2})$ values to $2_1^+$ and $2_2^+$
are consistent with this assignment.

\begin{table*}
\centering
\caption{\label{tab:E2_ZeN30}}
\begin{small}
$B({\rm E2})$ values ($e^2 {\rm fm}^4$) or E2 static moments
($e {\rm fm}^2$) in $Z={\em even}$, $N=30$ nuclei.
The experimental data are taken
from Refs.\cite{ref:NDS52,ref:NDS54,ref:NDS56,ref:NDS58}.\\
\begin{tabular}{|c||c|c||r@{.}l|r@{.}l@{$\pm$}r@{.}l|}
   \hline
     nucl. & $i$ & $f$ & \multicolumn{2}{c|}{Cal.}
      & \multicolumn{4}{c|}{Exp.} \\
   \hline
     $^{52}$Ti & $2_1^+$ & $0_1^+$ & $100$&$2~$
      & $138$&&\multicolumn{2}{l|}{$^{~104.}_{~~92.}$} \\
     & $2_2^+$ & $0_1^+$ & $13$&$7~$
      & $31$&&\multicolumn{2}{l|}{$^{~~23.}_{~~14.}$} \\
     & $2_3^+$ & $2_1^+$ & $53$&$7~$
      & \multicolumn{4}{l|}{$~~>~127.$} \\
     & $4_1^+$ & $2_1^+$ & $134$&$0~$
      & \multicolumn{4}{c|}{---} \\
     & $6_1^+$ & $4_1^+$ & $88$&$6~$ & $123$&&$~~22$& \\
     & $2_1^+$ & $2_1^+$ & $- 7$&$5^{*}$
      & \multicolumn{4}{c|}{---} \\
   \hline
     $^{54}$Cr & $2_1^+$ & $0_1^+$ & $179$&$1~$
      & $173$&$6$&$3$&$0$ \\
     & $4_1^+$ & $2_1^+$ & $237$&$6~$ & $303$&&$97$& \\
     & $2_2^+$ & $0_1^+$ & $1$&$2~$ & $10$&$9$&$4$&$8$ \\
     & $2_2^+$ & $2_1^+$ & $71$&$9~$ & $109$&&$36$& \\
     & $2_3^+$ & $2_1^+$ & $5$&$6~$
      & \multicolumn{4}{l|}{$~~>~291.$} \\
     & $6_1^+$ & $4_1^+$ & $196$&$8~$ & $218$&&$61$& \\
     & $2_1^+$ & $2_1^+$ & $- 24$&$5^{*}$
      & $-21$&&$8$&\multicolumn{1}{l|}{$^{*}$} \\
   \hline
     $^{56}$Fe & $2_1^+$ & $0_1^+$ & $164$&$8~$
      & $213$&$8$&$7$&$6$ \\
     & $4_1^+$ & $2_1^+$ & $229$&$6~$ & $305$&&$64$& \\
     & $2_2^+$ & $0_1^+$ & $2$&$7~$
      & $6$&$4$&\multicolumn{2}{l|}{$^{~~~7.6}_{~~~6.4}$} \\
     & $2_2^+$ & $2_1^+$ & $4$&$5~$ & $31$&&$14$& \\
     & $0_2^+$ & $2_1^+$ & $6$&$7~$ & $31$&&$15$& \\
     & $2_3^+$ & $0_1^+$ & $0$&$6~$ & $0$&$97$&$0$&$11$ \\
     & $2_3^+$ & $2_1^+$ & $13$&$7~$ & $16$&$5$&$5$&$1$ \\
     & $1_1^+$ & $2_1^+$ & $0$&$0004$ & $110$&$7$&$8$&$9$ \\
     & $4_2^+$ & $2_1^+$ & $0$&$0008$
      & $1$&$0$&\multicolumn{2}{l|}{$^{~~~1.1}_{~~~1.0}$} \\
     & $2_4^+$ & $0_1^+$ & $17$&$2~$ & $10$&$2$&$5$&$1$ \\
     & $2_4^+$ & $2_1^+$ & $18$&$0~$
      & $15$&&\multicolumn{2}{l|}{$^{~~19.}_{~~15.}$} \\
     & $6_1^+$ & $4_1^+$ & $33$&$4~$ & $50$&$9$&$5$&$1$ \\
   \hline
\end{tabular}
\begin{tabular}{|c||c|c||r@{.}l|r@{.}l@{$\pm$}r@{.}l|}
   \hline
     nucl. & $i$ & $f$ & \multicolumn{2}{c|}{Cal.}
      & \multicolumn{4}{c|}{Exp.} \\
   \hline
     $^{56}$Fe & $3_1^+$ & $2_1^+$ & $13$&$0~$
      & $7$&$6$&$~~~6$&$4$ \\
     & $3_1^+$ & $4_1^+$ & $5$&$7~$ & $10$&$2$&$3$&$8$ \\
     & $3_1^+$ & $2_2^+$ & $149$&$3~$ & $430$&&$140$& \\
     & $1_2^+$ & $2_1^+$ & $0$&$5~$ & $113$&&$28$& \\
     & $0_3^+$ & $2_1^+$ & $69$&$2~$
      & \multicolumn{4}{l|}{$~~>~~61.$} \\
     & $2_5^+$ & $0_1^+$ & $1$&$3~$ & $5$&$1$&$3$&$2$ \\
     & $2_5^+$ & $2_1^+$ & $0$&$002$ & $5$&$1$&$3$&$8$ \\
     & $6_2^+$ & $4_1^+$ & $193$&$6~$ & $267$&&$51$& \\
     & $6_2^+$ & $4_2^+$ & $2$&$1~$ & $380$&&$380$& \\
     & $6_2^+$ & $6_1^+$ & $81$&$9~$
      & $510$&&\multicolumn{2}{l|}{$^{~200.}_{~~50.}$} \\
     & $2_1^+$ & $2_1^+$ & $- 25$&$4^{*}$
      & $-23$&&$3$&\multicolumn{1}{l|}{$^{*}$} \\
   \hline
     $^{58}$Ni & $2_1^+$ & $0_1^+$ & $51$&$6~$ & $135$&$0$&$3$&$2$ \\
     & $4_1^+$ & $2_1^+$ & $38$&$4~$
      & \multicolumn{4}{l|}{$~~<~570.$} \\
     & $2_2^+$ & $0_1^+$ & $9$&$0~$ & $0$&$37$&$0$&$15$ \\
     & $2_2^+$ & $2_1^+$ & $21$&$2~$ & $200$&&$80$& \\
     & $0_2^+$ & $2_1^+$ & $22$&$4~$ & \multicolumn{4}{c|}{---} \\
     & $0_2^+$ & $2_2^+$ & $35$&$5~$ & \multicolumn{4}{c|}{---} \\
     & $0_3^+$ & $2_1^+$ & $0$&$3~$ & $0$&$0033$&$0$&$0001$ \\
     & $0_3^+$ & $2_2^+$ & $79$&$4~$ & $142$&$7$&$5$&$3$ \\
     & $2_3^+$ & $2_1^+$ & $3$&$0~$ & $35$&&$11$& \\
     & $2_3^+$ & $2_2^+$ & $3$&$1~$ & $130$&&$400$& \\
     & $2_3^+$ & $0_2^+$ & $199$&$9~$ & \multicolumn{4}{c|}{---} \\
     & $2_4^+$ & $0_1^+$ & $3$&$6~$ & $37$&$3$&$6$&$7$ \\
     & $2_4^+$ & $2_1^+$ & $18$&$4~$ & $130$&&$120$& \\
     & $3_1^+$ & $4_1^+$ & $0$&$008$ & $1$&$1$&$3$&$2$ \\
     & $2_1^+$ & $2_1^+$ & $- 11$&$3^{*}$
      & $-10$&&$6$&\multicolumn{1}{l|}{$^{*}$} \\
   \hline
\end{tabular}
\\
$^*$) Quadrupole moment.
\end{small}
\end{table*}

\clearpage
\subsection{$Z={\em odd},~N=30$ nuclei}
\label{subsec:E2_Zodd-N30}

In Table~\ref{tab:E2_ZoN30},
the calculated and measured E2 properties\footnote{
The $Q([{3\over 2}]_1^-)$ value of $^{57}$Co
quoted in Ref.\cite{ref:NDS57} is questionable,
since it is not even referred to
in the original paper\cite{ref:Co57QM}.}
are compared for $Z={\em odd}$, $N=30$ nuclei.

The agreement of the calculation with the data is good
in $^{53}$V and $^{55}$Mn.
Although the data in $^{57}$Co are not necessarily reproduced well,
we point out that the orders of magnitude are correct
for most transitions.

\begin{table*}
\centering
\caption{\label{tab:E2_ZoN30}}
\begin{small}
$B({\rm E2})$ values ($e^2 {\rm fm}^4$) or E2 static moments
($e {\rm fm}^2$) in $Z={\em odd}$, $N=30$ nuclei.
The experimental data are taken
from Refs.\cite{ref:NDS51,ref:NDS53,ref:NDS55,ref:NDS57}.\\
\begin{tabular}{|c||c|c||r@{.}l|r@{.}l@{$\pm$}r@{.}l|}
   \hline
     nucl. & $i$ & $f$ & \multicolumn{2}{c|}{Cal.}
      & \multicolumn{4}{c|}{Exp.} \\
   \hline
     $^{51}$Sc & $[{3\over 2}]_1^-$ & $[{7\over 2}]_1^-$
      & $50$&$0~$ & \multicolumn{4}{c|}{---} \\
     & $[{11\over 2}]_1^-$ & $[{7\over 2}]_1^-$
      & $22$&$7~$ & \multicolumn{4}{c|}{---} \\
     & $[{9\over 2}]_1^-$ & $[{7\over 2}]_1^-$
      & $28$&$2~$ & \multicolumn{4}{c|}{---} \\
     & $[{5\over 2}]_1^-$ & $[{7\over 2}]_1^-$
      & $3$&$4~$ & \multicolumn{4}{c|}{---} \\
     & $[{7\over 2}]_1^-$ & $[{7\over 2}]_1^-$
      & $- 23$&$8^{*}$ & \multicolumn{4}{c|}{---} \\
   \hline
     $^{53}$V & $[{5\over 2}]_1^-$ & $[{7\over 2}]_1^-$
      & $253$&$2~$ & \multicolumn{4}{c|}{---} \\
     & $[{3\over 2}]_1^-$ & $[{7\over 2}]_1^-$
      & $162$&$2~$ & $148$&&$~~12$& \\
     & $[{11\over 2}]_1^-$ & $[{7\over 2}]_1^-$
      & $153$&$3~$ & $183$&&$28$& \\
     & $[{15\over 2}]_1^-$ & $[{11\over 2}]_1^-$
      & $157$&$7~$ & $154$&&$35$& \\
     & $[{7\over 2}]_1^-$ & $[{7\over 2}]_1^-$
      & $- 10$&$7^{*}$ & \multicolumn{4}{c|}{---} \\
   \hline
     $^{55}$Mn & $[{7\over 2}]_1^-$ & $[{5\over 2}]_1^-$
      & $287$&$2~$ & $335$&&$99$& \\
     & $[{9\over 2}]_1^-$ & $[{5\over 2}]_1^-$
      & $85$&$8~$ & $109$&&$14$& \\
     & $[{9\over 2}]_1^-$ & $[{7\over 2}]_1^-$
      & $195$&$6~$ & $286$&&$37$& \\
     & $[{11\over 2}]_1^-$ & $[{7\over 2}]_1^-$
      & $152$&$6~$ & $175$&&$15$& \\
     & $[{11\over 2}]_1^-$ & $[{9\over 2}]_1^-$
      & $160$&$3~$ & \multicolumn{4}{l|}{$~~<~~87.$} \\
     & $[{1\over 2}]_1^-$ & $[{5\over 2}]_1^-$
      & $36$&$2~$ & $144$&&$12$& \\
     & $[{3\over 2}]_1^-$ & $[{5\over 2}]_1^-$
      & $19$&$6~$ & $41$&&$26$& \\
     & $[{13\over 2}]_1^-$ & $[{9\over 2}]_1^-$
      & $148$&$5~$ & $99$&&$50$& \\
     & $[{13\over 2}]_1^-$ & $[{11\over 2}]_1^-$
      & $89$&$9~$ & $40$&&$11$& \\
     & $[{15\over 2}]_1^-$ & $[{11\over 2}]_1^-$
      & $164$&$2~$ & $96$&&$22$& \\
     & $[{15\over 2}]_1^-$ & $[{13\over 2}]_1^-$
      & $52$&$0~$ & $32$&&$20$& \\
     & $[{5\over 2}]_1^-$ & $[{5\over 2}]_1^-$
      & $33$&$5^{*}$ & $33$&&$1$&\multicolumn{1}{l|}{$^{*}$} \\
   \hline
\end{tabular}
\begin{tabular}{|c||c|c||r@{.}l|r@{.}l@{$\pm$}r@{.}l|}
   \hline
     nucl. & $i$ & $f$ & \multicolumn{2}{c|}{Cal.}
      & \multicolumn{4}{c|}{Exp.} \\
\hline
     $^{57}$Co & $[{9\over 2}]_1^-$ & $[{7\over 2}]_1^-$
      & $135$&$3~$ & $241$&&$~~29$& \\
     & $[{3\over 2}]_1^-$ & $[{7\over 2}]_1^-$
      & $0$&$08$ & $6$&$0$&$1$&$3$ \\
     & $[{1\over 2}]_1^-$ & $[{3\over 2}]_1^-$
      & $480$&$8~$ & \multicolumn{4}{l|}{$~~<~~23.$} \\
     & $[{11\over 2}]_1^-$ & $[{7\over 2}]_1^-$
      & $34$&$1~$ & $78$&$2$&$7$&$8$ \\
     & $[{11\over 2}]_1^-$ & $[{9\over 2}]_1^-$
      & $114$&$9~$ & $370$&&$100$& \\
     & $[{7\over 2}]_2^-$ & $[{7\over 2}]_1^-$
      & $4$&$9~$ & $0$&$1$&$1$&$6$ \\
     & $[{7\over 2}]_2^-$ & $[{9\over 2}]_1^-$
      & $2$&$8~$ & $9$&$1$&$9$&$1$ \\
     & $[{7\over 2}]_2^-$ & $[{3\over 2}]_1^-$
      & $254$&$7~$ & \multicolumn{4}{c|}{---} \\
     & $[{5\over 2}]_1^-$ & $[{7\over 2}]_1^-$
      & $4$&$5~$ & $48$&&$14$& \\
     & $[{5\over 2}]_2^-$ & $[{7\over 2}]_1^-$
      & $8$&$7~$ & \multicolumn{4}{c|}{---} \\
     & $[{7\over 2}]_3^-$ & $[{7\over 2}]_1^-$
      & $8$&$7~$ & \multicolumn{4}{l|}{$~~<~~~4.2$} \\
     & $[{7\over 2}]_3^-$ & $[{9\over 2}]_1^-$
      & $31$&$6~$ & $20$&$8$&$7$&$8$ \\
     & $[{7\over 2}]_3^-$ & $[{3\over 2}]_1^-$
      & $42$&$7~$ & $370$&&$120$& \\
     & $[{9\over 2}]_2^-$ & $[{7\over 2}]_1^-$
      & $1$&$1~$ & \multicolumn{4}{l|}{$~~<~~78.$} \\
     & $[{9\over 2}]_2^-$ & $[{9\over 2}]_1^-$
      & $0$&$3~$ & \multicolumn{4}{l|}{$~~<~500.$} \\
     & $[{9\over 2}]_2^-$ & $[{11\over 2}]_1^-$
      & $18$&$8~$ & \multicolumn{4}{l|}{$~~<5000.$} \\
     & $[{13\over 2}]_1^-$ & $[{11\over 2}]_1^-$
      & $101$&$8~$ & $340$&&$330$& \\
     & $[{7\over 2}]_1^-$ & $[{7\over 2}]_1^-$
      & $32$&$7^{*}$ & $52$&&$9$&\multicolumn{1}{l|}{$^{*}$} \\
     & $[{3\over 2}]_1^-$ & $[{3\over 2}]_1^-$
      & $- 21$&$8^{*}$ & \multicolumn{4}{c|}{---} \\
   \hline
\end{tabular}
\\
$^*$) Quadrupole moment.
\end{small}
\end{table*}

\clearpage
\section{M1 properties}
\label{sec:M1}

\subsection{Single-particle parameters}
\label{subsec:M1sp}

The M1 properties are usually described
by the following one-body operator,
\be T({\rm M1}) = \sqrt{3\over{4\pi}} \sum_i \hat\mu_i .
 \label{eq4:M1sp} \ee
For the free nucleon, we have
\be \hat\mu_i = \left\{ \begin{array}{ll}
     \hat l_i + g^{\rm free}_{s, \pi} \hat s_i &
     ~(i\in \mbox{proton}) \\
     g^{\rm free}_{s, \nu} \hat s_i &~(i\in \mbox{neutron})
      \end{array} \right. ~, \ee
in units of $\mu_{\rm N}$,
where $l_i$ ($s_i$) denotes the orbital (spin) angular momentum
of the $i$-th nucleon,
and the $g_s$'s are spin $g$-factors
with $g^{\rm free}_{s, \pi}=5.58$, $g^{\rm free}_{s, \nu}=-3.82$.
It is known, however, that the core polarization
and the meson exchange current
contribute to the M1 properties.
The former appears to be the price for the truncation
of the model space in the shell model calculation.
Especially the quenching of the spin $g$-factor
is required in most cases.
Considering the orbit dependence of the single-particle parameters
as in the E2 case,
we can parametrize the M1 operator as
\be \hat\mu_i = g_{l,\rho}^{\rm eff}(nl) \hat l_i
 + g_{s,\rho}^{\rm eff}(nl) \hat s_i
+ g_{p,\rho}^{\rm eff}(nl, n'l') [Y^{(2)}(\hat{\bm{r}}_i) s_i]^{(1)}
{}~(\rho =\pi, \nu) , \label{eq4:M1op0} \ee
where the $g_l$'s are orbital $g$-factors.
The $g$-factors are, in principle,
dependent on $n$ and $l$ quantum numbers
of the single-particle orbit.
We adopt the microscopic parameters
that Towner calculated with the single-particle wavefunctions
in the harmonic oscillator approximation\cite{ref:Towner}.

As in the calculation of energy levels and E2 properties,
the mass-number dependence
of the single-particle parameters is neglected.
In the harmonic oscillator approximation,
only the radial part varies with mass number.
The matrix elements of the $\hat l$ and $\hat s$ operators,
which form the bare M1 operator,
do not depend on the radial part.
Only a part of the correction coming from the core polarization
and meson exchange effects
will change with $A$.
Therefore the $A$-dependence of the M1 single-particle parameters
will be weak.

\clearpage
\subsection{$Z={\em even},~N=28$ nuclei}
\label{subsec:M1_Zeven-N28}

The calculated M1 properties
are shown in Table~\ref{tab:M1_ZeN28},
in comparison with the experimental data,
for $Z={\em even}$, $N=28$ nuclei.

All the measured magnetic moments are reproduced very well
in $^{50}$Ti, $^{52}$Cr and $^{54}$Fe.
In the $k=0$ configuration,
the particle-hole conjugation leads to the same $\mu (2_1^+)$
between $^{50}$Ti and $^{54}$Fe.
This situation approximately holds both in the experiment
and in the $k\leq 2$ calculation.
The contributions of the higher configurations,
which have been suggested with respect to the E2 quantities,
are not clear in the magnetic moments.
The $\mu (2_1^+)$ value of $^{54}$Cr is close to
those of $^{50}$Ti and $^{54}$Fe,
as is expected if the seniority conservation is satisfied
in the $k=0$ space.
It should be mentioned that a smaller experimental value
of $\mu(2_1^+)=2.00\pm 0.32[\mu_{\rm N}^2]$
has been indicated for $^{54}$Fe recently\cite{ref:Fe54MM}.

In $^{52}$Cr,
the $B({\rm M1})$ values from $4_3^+$ imply
that the $4_1^+$ and $4_2^+$ assignment performed
in Section~\ref{sec:E2} is reasonable.

\begin{table*}
\centering
\caption{\label{tab:M1_ZeN28}}
\begin{small}
$B({\rm M1})$ values ($\mu_{\rm N}^2$) or M1 static moments
($\mu_{\rm N}$) in $Z={\em even}$, $N=28$ nuclei.
The `Cal.' values are obtained by the present calculation.
The experimental data (Exp.) are taken
from Refs.\cite{ref:NDS50,ref:NDS52,ref:NDS54}.\\
\begin{tabular}{|c||c|c||r@{.}l|r@{.}l@{$\pm$}r@{.}l|}
   \hline
     nucl. & $i$ & $f$ & \multicolumn{2}{c|}{Cal.}
      & \multicolumn{4}{c|}{Exp.} \\
   \hline
     $^{50}$Ti & $2_1^+$ & $2_1^+$ & $2$&$388^{*}$
      & $2$&$4$&$~0$&$8~^{*}$ \\
   \hline
     $^{52}$Cr & $2_2^+$ & $2_1^+$ & $0$&$001$
      & $0$&$0006$&$0$&$0001$ \\
     & $2_3^+$ & $2_1^+$ & $0$&$009$ & $0$&$082$&$0$&$027$ \\
     & $4_3^+$ & $4_1^+$ & $0$&$920$ & \multicolumn{4}{c|}{---} \\
     & $4_3^+$ & $4_2^+$ & $0$&$216$ & $0$&$358$&$0$&$072$ \\
   \hline
\end{tabular}
\begin{tabular}{|c||c|c||r@{.}l|r@{.}l@{$\pm$}r@{.}l|}
   \hline
     nucl. & $i$ & $f$ & \multicolumn{2}{c|}{Cal.}
      & \multicolumn{4}{c|}{Exp.} \\
   \hline
     $^{52}$Cr & $3_1^+$ & $4_2^+$ & $0$&$009$
      & $0$&$013$&$0$&$005$ \\
     & $2_1^+$ & $2_1^+$ & $2$&$732^{*}$ & $3$&$00$&$0$&$50^{*}$ \\
   \hline
     $^{54}$Fe & $2_1^+$ & $2_1^+$ & $2$&$732^{*}$
      & $2$&$4$&$0$&$3~^{*}$ \\
     & $6_1^+$ & $6_1^+$ & $8$&$334^{*}$ & $8$&$22$&$0$&$18^{*}$ \\
   \hline
\end{tabular}
\\
$^*$) Magnetic moment.
\end{small}
\end{table*}

\clearpage
\subsection{$Z={\em odd},~N=28$ nuclei}
\label{subsec:M1_Zodd-N28}

The M1 quantities in $Z={\em odd}$, $N=28$ nuclei
are displayed in Table~\ref{tab:M1_ZoN28}.

All the measured magnetic moments are reproduced remarkably well,
highlightening the adequacy of the present description
with respect to the proton degrees of freedom.
According to the particle-hole symmetry in the $k=0$ space,
the same magnetic moments are expected between $^{51}$V and $^{53}$Mn,
as well as between $^{49}$Sc and $^{55}$Co.
This equality is satisfied approximately.

As in the case of $^{52}$Cr, the qualitative trend
of the $B({\rm M1})$ values
is reproduced in $^{51}$V and $^{53}$Mn.
{}From a quantitative point of view,
we point out that the $B({\rm M1})$ values are generally underestimated
in the $Z={\em odd}$ or ${\em even}$, $N=28$ nuclei.

\begin{table*}
\centering
\caption{\label{tab:M1_ZoN28}}
\begin{small}
$B({\rm M1})$ values ($\mu_{\rm N}^2$) or M1 static moments
($\mu_{\rm N}$) in $Z={\em odd}$, $N=28$ nuclei.
The experimental data are taken
from Refs.\cite{ref:NDS49,ref:NDS51,ref:NDS53,ref:NDS55}.\\
\begin{tabular}{|c||c|c||r@{.}l|r@{.}l@{$\pm$}r@{.}l|}
   \hline
     nucl. & $i$ & $f$ & \multicolumn{2}{c|}{Cal.}
      & \multicolumn{4}{c|}{Exp.} \\
   \hline
     $^{49}$Sc & $[{7\over 2}]_1^-$ & $[{7\over 2}]_1^-$
      & $5$&$152^{*}$ & \multicolumn{4}{c|}{---} \\
   \hline
     $^{51}$V & $[{5\over 2}]_1^-$ & $[{7\over 2}]_1^-$
      & $0$&$0003$ & $0$&$0053$&$~0$&$0003$ \\
     & $[{3\over 2}]_1^-$ & $[{5\over 2}]_1^-$
      & $0$&$0008$ & $0$&$00007$&$0$&$00002$ \\
     & $[{9\over 2}]_1^-$ & $[{7\over 2}]_1^-$
      & $0$&$0005$ & $0$&$0006$&$0$&$0002$ \\
     & $[{9\over 2}]_1^-$ & $[{11\over 2}]_1^-$
      & $0$&$033$ & $0$&$082$&$0$&$045$ \\
     & $[{7\over 2}]_1^-$ & $[{7\over 2}]_1^-$
      & $5$&$008^{*}$ & $5$&$149$&$0$&$000^{*}$ \\
     & $[{5\over 2}]_1^-$ & $[{5\over 2}]_1^-$
      & $3$&$457^{*}$ & $3$&$86$&$0$&$33~^{*}$ \\
   \hline
\end{tabular}
\begin{tabular}{|c||c|c||r@{.}l|r@{.}l@{$\pm$}r@{.}l|}
   \hline
     nucl. & $i$ & $f$ & \multicolumn{2}{c|}{Cal.}
      & \multicolumn{4}{c|}{Exp.} \\
   \hline
     $^{53}$Mn & $[{5\over 2}]_1^-$ & $[{7\over 2}]_1^-$
      & $0$&$0001$ & $0$&$0046$&$0$&$0004$ \\
     & $[{3\over 2}]_1^-$ & $[{5\over 2}]_1^-$
      & $0$&$0009$ & $0$&$041$&$0$&$005$ \\
     & $[{9\over 2}]_1^-$ & $[{7\over 2}]_1^-$
      & $0$&$00006$ & $0$&$0021$&$0$&$0005$ \\
     & $[{13\over 2}]_1^-$ & $[{11\over 2}]_1^-$
      & $0$&$0009$ & $0$&$0026$&$0$&$0003$ \\
     & $[{15\over 2}]_1^-$ & $[{13\over 2}]_1^-$
      & $0$&$024$ & $0$&$376$&$0$&$072$ \\
     & $[{7\over 2}]_1^-$ & $[{7\over 2}]_1^-$
      & $4$&$996^{*}$ & $5$&$024$&$0$&$007^{*}$ \\
     & $[{5\over 2}]_1^-$ & $[{5\over 2}]_1^-$
      & $3$&$488^{*}$ & $3$&$25$&$0$&$30~^{*}$ \\
   \hline
     $^{55}$Co & $[{7\over 2}]_1^-$ & $[{7\over 2}]_1^-$
      & $4$&$966^{*}$ & $4$&$822$&$0$&$003^{*}$ \\
   \hline
\end{tabular}
\\
$^*$) Magnetic moment.
\end{small}
\end{table*}

\clearpage
\subsection{$Z={\em even},~N=29$ nuclei}
\label{subsec:M1_Zeven-N29}

Table~\ref{tab:M1_ZeN29} shows the M1 properties
for $Z={\em even}$, $N=29$ nuclei.

We first look at $^{57}$Ni,
which could be a good test of the neutron degrees of freedom.
In Table~\ref{tab:M1_ZeN29},
the sign of the experimental magnetic moment of $[{3\over 2}]_1^-$,
which has not been specified\cite{ref:Ni57MM},
is conjectured from the calculated value.
The adopted sign is consistent with the Schmidt value.
The $\mu ([{3\over 2}]_1^-)$
and $B({\rm M1}; [{1\over 2}]_1^- \rightarrow [{3\over 2}]_1^-)$ values
are overestimated in the present calculation.
Since the neutron orbital angular momentum hardly
affects the M1 observables,
it is suggested that further quenching of $g_{s,\nu}$
is required for more precise description.
Though within the $k=0$ configuration
the M1 transition from $[{5\over 2}]_1^-$ to $[{3\over 2}]_1^-$
is possible only through the $[Y^{(2)}s]^{(1)}$-term,
the measured probability is significantly larger
than expected from a reasonable $g_{p,\nu}$ value.
Analogously, the present calculation underestimates
$B({\rm M1}; [{5\over 2}]_1^- \rightarrow [{3\over 2}]_1^-)$
by several orders of magnitude.
In order to remedy this problem,
we should include higher configuration into the wavefunctions,
or two-body terms into the M1 operator.

It should be noticed that the calculated $\mu ([{7\over 2}]^-_1)$
and $\mu ([{7\over 2}]^-_2)$ of $^{53}$Cr
have opposite signs to each other.
The observed $\mu ([{7\over 2}]^-_1)$ prefers to be positive.
This fact is consistent with the present assignment
based on the $B({\rm E2})$ values in Section~\ref{sec:E2},
in which the indices of the calculated two ${7\over 2}^-$ states
have been inverted
against their energy sequence.
The signs of the M1 moments of the lowest two ${7\over 2}^-$ states
are correctly reproduced also in $^{55}$Fe,
even though the absolute value of $\mu ([{7\over 2}]^-_2)$
is underestimated to a certain extent.

The calculated M1 transition rates agree with the data,
at least within the order of magnitude,
for $^{51}$Ti, $^{53}$Cr and $^{55}$Fe.
The only exception is
$B({\rm M1}; [{7\over 2}]^-_2 \rightarrow [{5\over 2}]^-_1)$ in $^{55}$Fe,
which is not a serious problem
since this $B({\rm M1})$ value is small
both theoretically and experimentally.

\begin{table*}
\centering
\caption{\label{tab:M1_ZeN29}}
\begin{small}
$B({\rm M1})$ values ($\mu_{\rm N}^2$) or M1 static moments
($\mu_{\rm N}$) in $Z={\em even}$, $N=29$ nuclei.
The experimental data are taken
from Refs.\cite{ref:NDS51,ref:NDS53,ref:NDS55,ref:NDS57}.\\
\begin{tabular}{|c||c|c||r@{.}l|r@{.}l@{$\pm$}r@{.}l|}
   \hline
     nucl. & $i$ & $f$ & \multicolumn{2}{c|}{Cal.}
      & \multicolumn{4}{c|}{Exp.} \\
   \hline
     $^{51}$Ti & $[{1\over 2}]_1^-$ & $[{3\over 2}]_1^-$
      & $0$&$825$ & $0$&$734$&$~0$&$090$ \\
     & $[{5\over 2}]_1^-$ & $[{3\over 2}]_1^-$
      & $0$&$049$ & $0$&$059$&$0$&$018$ \\
     & $[{5\over 2}]_2^-$ & $[{3\over 2}]_1^-$
      & $0$&$002$ & $0$&$016$&$0$&$003$ \\
     & $[{5\over 2}]_2^-$ & $[{7\over 2}]_1^-$
      & $0$&$068$ & $0$&$013$&$0$&$003$ \\
     & $[{5\over 2}]_2^-$ & $[{5\over 2}]_1^-$
      & $0$&$146$ & $0$&$170$&$0$&$032$ \\
     & $[{3\over 2}]_2^-$ & $[{3\over 2}]_1^-$
      & $0$&$085$ & $0$&$127$&$0$&$004$ \\
     & $[{3\over 2}]_2^-$ & $[{1\over 2}]_1^-$
      & $0$&$146$ & $0$&$516$&$0$&$016$ \\
     & $[{3\over 2}]_2^-$ & $[{5\over 2}]_1^-$
      & $1$&$739$ & $3$&$22$&$0$&$11$ \\
     & $[{3\over 2}]_1^-$ & $[{3\over 2}]_1^-$
      & $-0$&$989^{*}$ & \multicolumn{4}{c|}{---} \\
   \hline
     $^{53}$Cr & $[{1\over 2}]_1^-$ & $[{3\over 2}]_1^-$
      & $0$&$653$ & \multicolumn{4}{c|}{---} \\
     & $[{5\over 2}]_1^-$ & $[{3\over 2}]_1^-$
      & $0$&$003$ & $0$&$093$&$0$&$013$ \\
     & $[{7\over 2}]_1^-$ & $[{5\over 2}]_1^-$
      & $0$&$273$ & $0$&$082$&$0$&$016$ \\
     & $[{7\over 2}]_2^-$ & $[{5\over 2}]_1^-$
      & $0$&$001$ & $0$&$0075$&$0$&$0005$ \\
     & $[{7\over 2}]_2^-$ & $[{7\over 2}]_1^-$
      & $0$&$00006$ & $0$&$029$&$0$&$003$ \\
     & $[{5\over 2}]_2^-$ & $[{3\over 2}]_1^-$
      & $0$&$076$ & $0$&$050$&$0$&$016$ \\
     & $[{5\over 2}]_2^-$ & $[{7\over 2}]_1^-$
      & $0$&$310$ & $0$&$269$&$0$&$090$ \\
     & $[{5\over 2}]_2^-$ & $[{7\over 2}]_2^-$
      & $0$&$00000$ & \multicolumn{4}{c|}{---} \\
     & $[{9\over 2}]_1^-$ & $[{7\over 2}]_1^-$
      & $0$&$001$ & \multicolumn{4}{c|}{---} \\
     & $[{9\over 2}]_1^-$ & $[{7\over 2}]_2^-$
      & $0$&$618$ & $0$&$315$&$0$&$045$ \\
   \hline
\end{tabular}
\begin{tabular}{|c||c|c||r@{.}l|r@{.}l@{$\pm$}r@{.}l|}
   \hline
     nucl. & $i$ & $f$ & \multicolumn{2}{c|}{Cal.}
      &  \multicolumn{4}{c|}{Exp.} \\
   \hline
     $^{53}$Cr & $[{3\over 2}]_2^-$ & $[{3\over 2}]_1^-$
      & $0$&$533$ & $3$&$6$&$7$&$2$ \\
     & $[{3\over 2}]_1^-$ & $[{3\over 2}]_1^-$
      & $-0$&$442^{*}$ & $-0$&$475$&\multicolumn{2}{l|}{$~~~^{*}$} \\
     & $[{7\over 2}]_1^-$ & $[{7\over 2}]_1^-$
      & $2$&$435^{*}$ & $2$&$8$&$4$&$9~^{*}$ \\
     & $[{7\over 2}]_2^-$ & $[{7\over 2}]_2^-$
      & $-0$&$719^{*}$ & \multicolumn{4}{c|}{---} \\
   \hline
     $^{55}$Fe & $[{1\over 2}]_1^-$ & $[{3\over 2}]_1^-$
      & $0$&$844$ & \multicolumn{4}{c|}{---} \\
     & $[{5\over 2}]_1^-$ & $[{3\over 2}]_1^-$
      & $0$&$0004$ & $0$&$005$&$0$&$002$ \\
     & $[{7\over 2}]_1^-$ & $[{5\over 2}]_1^-$
      & $0$&$123$ & $0$&$023$&$0$&$016$ \\
     & $[{7\over 2}]_2^-$ & $[{5\over 2}]_1^-$
      & $0$&$00007$ & $0$&$0050$&$0$&$0004$ \\
     & $[{9\over 2}]_1^-$ & $[{7\over 2}]_2^-$
      & $0$&$317$ & $0$&$093$&$0$&$027$ \\
     & $[{9\over 2}]_2^-$ & $[{7\over 2}]_1^-$
      & $0$&$001$ & $0$&$005$&$0$&$005$ \\
     & $[{3\over 2}]_1^-$ & $[{3\over 2}]_1^-$
      & $-0$&$616^{*}$ & \multicolumn{4}{c|}{---} \\
     & $[{5\over 2}]_1^-$ & $[{5\over 2}]_1^-$
      & $1$&$446^{*}$ & $2$&$7$&$1$&$2~^{*}$ \\
     & $[{7\over 2}]_1^-$ & $[{7\over 2}]_1^-$
      & $2$&$133^{*}$ & $2$&&$2$&\multicolumn{1}{l|}{$~^{*}$} \\
     & $[{7\over 2}]_2^-$ & $[{7\over 2}]_2^-$
      & $-0$&$764^{*}$ & $-2$&$2$&$0$&$5~^{*}$ \\
   \hline
     $^{57}$Ni & $[{5\over 2}]_1^-$ & $[{3\over 2}]_1^-$
      & $0$&$00001$ & $0$&$026$&$0$&$003$ \\
     & $[{1\over 2}]_1^-$ & $[{3\over 2}]_1^-$
      & $0$&$675$ & \multicolumn{4}{l|}{$~~<~0.34$} \\
     & $[{3\over 2}]_1^-$ & $[{3\over 2}]_1^-$
      & $-1$&$108^{*}$ & $-0$&$88$&$0$&$06^{*}$ \\
   \hline
\end{tabular}
\\
$^*$) Magnetic moment.
\end{small}
\end{table*}

\clearpage
\subsection{$Z={\em odd},~N=29$ nuclei}
\label{subsec:M1_Zodd-N29}

Table~\ref{tab:M1_ZoN29} shows the M1 properties
for $Z={\em odd}$, $N=29$ nuclei.
We obtain good agreement,
except for the extremely large $\mu (4_1^+)$
and $\mu (5_1^+)$ of $^{54}$Mn
which are beyond the ordinary shell model description.

\begin{table*}
\centering
\caption{\label{tab:M1_ZoN29}}
\begin{small}
$B({\rm M1})$ values ($\mu_{\rm N}^2$) or M1 static moments
($\mu_{\rm N}$) in $Z={\em odd}$, $N=29$ nuclei.
The experimental data are taken
from Refs.\cite{ref:NDS50,ref:NDS52,ref:NDS54,ref:NDS56}.\\
\begin{tabular}{|c||c|c||r@{.}l|r@{.}l@{$\pm$}r@{.}l|}
   \hline
     nucl. & $i$ & $f$ & \multicolumn{2}{c|}{Cal.}
      & \multicolumn{4}{c|}{Exp.} \\
   \hline
     $^{50}$Sc & $3_1^+$ & $2_1^+$ & $2$&$099$
      & \multicolumn{4}{c|}{---} \\
     & $4_1^+$ & $5_1^+$ & $1$&$748$ & \multicolumn{4}{c|}{---} \\
     & $5_1^+$ & $5_1^+$ & $3$&$980^{*}$ & \multicolumn{4}{c|}{---} \\
   \hline
     $^{52}$V & $2_1^+$ & $3_1^+$ & $1$&$528$ & $1$&$33$&$~0$&$50$ \\
     & $1_1^+$ & $2_1^+$ & $0$&$226$ & \multicolumn{4}{c|}{---} \\
     & $4_1^+$ & $3_1^+$ & $0$&$110$ & \multicolumn{4}{c|}{---} \\
     & $4_1^+$ & $5_1^+$ & $0$&$572$ & \multicolumn{4}{c|}{---} \\
     & $3_1^+$ & $3_1^+$ & $3$&$049^{*}$ & \multicolumn{4}{c|}{---} \\
   \hline
     $^{54}$Mn & $2_1^+$ & $3_1^+$ & $1$&$158$
      & \multicolumn{4}{c|}{---} \\
     & $4_1^+$ & $3_1^+$ & $0$&$100$ & $0$&$055$&$0$&$005$ \\
     & $5_1^+$ & $4_1^+$ & $0$&$538$ & $0$&$591$&$0$&$072$ \\
     & $3_2^+$ & $3_1^+$ & $0$&$099$ & $0$&$116$&$0$&$025$ \\
     & $3_2^+$ & $2_1^+$ & $0$&$034$ & $0$&$052$&$0$&$013$ \\
     & $3_2^+$ & $4_1^+$ & $1$&$182$ & $1$&$00$&$0$&$22$ \\
     & $4_2^+$ & $5_1^+$ & $0$&$499$ & $0$&$322$&$0$&$072$ \\
     & $3_3^+$ & $2_1^+$ & $0$&$420$ & $0$&$59$&$0$&$14$ \\
     & $3_3^+$ & $4_1^+$ & $0$&$245$ & $0$&$63$&$0$&$14$ \\
     & $6_1^+$ & $5_1^+$ & $0$&$0004$ & $0$&$0005$&$0$&$0002$ \\
   \hline
\end{tabular}
\begin{tabular}{|c||c|c||r@{.}l|r@{.}l@{$\pm$}r@{.}l|}
   \hline
     nucl. & $i$ & $f$ & \multicolumn{2}{c|}{Cal.}
      &  \multicolumn{4}{c|}{Exp.} \\
   \hline
     $^{54}$Mn & $1_1^+$ & $2_1^+$ & $0$&$030$
      & $0$&$025$&$0$&$014$ \\
     & $3_1^+$ & $3_1^+$ & $2$&$987^{*}$
      & $3$&$282$&$0$&$001^{*}$ \\
     & $4_1^+$ & $4_1^+$ & $3$&$749^{*}$ & $7$&$3$&$1$&$4~^{*}$ \\
     & $5_1^+$ & $5_1^+$ & $4$&$282^{*}$
      & $55$&&$30$&\multicolumn{1}{l|}{$~^{*}$} \\
     & $3_2^+$ & $3_2^+$ & $4$&$226^{*}$
      & \multicolumn{4}{l|}{$~~<18$} \\
   \hline
     $^{56}$Co & $3_1^+$ & $4_1^+$ & $2$&$190$
      & \multicolumn{4}{l|}{$~~>~0.10$} \\
     & $5_1^+$ & $4_1^+$ & $0$&$552$ & $0$&$72$&$0$&$18$ \\
     & $4_2^+$ & $4_1^+$ & $0$&$003$
      & \multicolumn{4}{l|}{$~~<~0.010$} \\
     & $4_2^+$ & $3_1^+$ & $0$&$035$
      & \multicolumn{4}{l|}{$~~<~0.057$} \\
     & $4_2^+$ & $5_1^+$ & $0$&$003$
      & \multicolumn{4}{l|}{$~~<~0.021$} \\
     & $2_1^+$ & $3_1^+$ & $0$&$855$ & $0$&$97$&$0$&$27$ \\
     & $5_2^+$ & $4_1^+$ & $0$&$223$ & $0$&$088$&$0$&$036$ \\
     & $3_2^+$ & $4_1^+$ & $0$&$103$ & $0$&$125$&$0$&$072$ \\
     & $3_2^+$ & $3_1^+$ & $0$&$006$ & $0$&$006$&$0$&$003$ \\
     & $3_2^+$ & $4_2^+$ & $0$&$727$ & $1$&$07$&$0$&$54$ \\
     & $1_1^+$ & $2_1^+$ & $0$&$00006$ & $0$&$143$&$0$&$072$ \\
     & $6_1^+$ & $5_1^+$ & $0$&$142$ & $0$&$125$&$0$&$072$ \\
     & $4_1^+$ & $4_1^+$ & $3$&$654^{*}$
      & $3$&$830$&$0$&$015^{*}$ \\
   \hline
\end{tabular}
\\
$^*$) Magnetic moment.
\end{small}
\end{table*}

\clearpage
\subsection{$Z={\em even},~N=30$ nuclei}
\label{subsec:M1_Zeven-N30}

The calculated M1 quantities are compared with the measured ones
in Table~\ref{tab:M1_ZeN30}, for $Z={\em even}$, $N=30$ nuclei.

Although the agreement of the $B({\rm M1})$ values
is not necessarily good in $^{56}$Fe,
the orders of magnitudes are correctly reproduced
for most transitions.
As will be discussed below,
the further quenching of $g_{s,\rho}$
improves the results considerably.
The observed $B({\rm M1}; 1_1^+ \rightarrow 0_1^+)$ value
is quite small.
This nature is reproduced well in the present calculation.
The discrepancy in $B({\rm M1}; 2_3^+ \rightarrow 2_1^+)$
is not improved by the $g_s$ quenching.
The cancellation among the elements of the density matrix
brings this small $B({\rm M1})$ value.
As discussed in Ref.\cite{ref:Nakada},
the $2_3^+$ state is a fully non-collective state.
We should note that non-collective transitions
are generally sensitive to details of wavefunctions.
The $(p, p')$ result\cite{ref:Takamatsu}
confirms the adequacy of the present shell model wavefunction
of this state,
as will be mentioned in Section~\ref{sec:pp'}.

In Ref.\cite{ref:Nakada},
the $n$ and $l$ dependences in Eq.(\ref{eq4:M1op0}) were neglected,
as is often assumed.
Furthermore, $g_{p, \rho}^{\rm eff} =0$ was assumed.
Namely, the employed M1 operator was
\be T({\rm M1}) = \sqrt{3\over{4\pi}} \sum_{\rho=\pi,\nu}
 ( g_{l, \rho}^{\rm eff} \hat L_\rho
  + g_{s, \rho}^{\rm eff} \hat S_\rho ) , \label{eq4:M1op} \ee
where
\be \hat L_\rho = \sum_{i\in \rho} \hat l_i,
{}~\hat S_\rho = \sum_{i\in \rho} \hat s_i . \label{eq:L&S} \ee
The single-particle $g$-factors of
$g_{l, \pi}^{\rm eff} = 1.0$, $g_{l, \nu}^{\rm eff} = 0.0$,
$g_{s, \pi}^{\rm eff} = 0.5 g_{s,\pi}^{\rm free}$
and $g_{s, \nu}^{\rm eff} = 0.5 g_{s,\nu}^{\rm free}$
were adopted in Ref.\cite{ref:Nakada}
so as to fit the data in $^{56}$Fe,
though microscopic calculations predict
the quenching factor for $g_{s,\rho}$
to be much closer to the unity\cite{ref:Towner,ref:M1review}.
It was shown in Ref.\cite{ref:Nakada}
that by this phenomenological set of parameters
the $B({\rm M1})$ valued in $^{56}$Fe are reproduced
reasonably well.

It is found that the predicted $1_3^+$ state of $^{56}$Fe,
as well as the $1_2^+$ state,
has a relatively large $B({\rm M1})$ value to the ground state.
This fact suggests
that these two states contain a considerable fraction
of a mixed-symmetry $1^+$ component.
A recent experiment has also reported
a mixed-symmetry $1^+$ strength
around $Ex\simeq 3.5$MeV\cite{ref:Fe56MS1}.
We will discuss this point in more detail in a forthcoming paper.
These two states are almost degenerate in energy ($Ex\simeq 3.5$MeV)
according to the calculation,
while only a single state has been
experimentally confirmed\cite{ref:NDS56}.
It is desired to search for the other state.

We do not have a good agreement for $^{58}$Ni.
This disagreement might imply a non-negligible influence
of the $k>2$ configurations.

\begin{table*}
\centering
\caption{\label{tab:M1_ZeN30}}
\begin{small}
$B({\rm M1})$ values ($\mu_{\rm N}^2$) or M1 static moments
($\mu_{\rm N}$) in $Z={\em even}$, $N=30$ nuclei.
The experimental data are taken
from Refs.\cite{ref:NDS52,ref:NDS54,ref:NDS56,ref:NDS58}.\\
\begin{tabular}{|c||c|c||r@{.}l|r@{.}l@{$\pm$}r@{.}l|}
   \hline
     nucl. & $i$ & $f$ & \multicolumn{2}{c|}{Cal.}
      & \multicolumn{4}{c|}{Exp.} \\
   \hline
     $^{52}$Ti & $2_2^+$ & $2_1^+$ & $0$&$456$
      & $0$&$56$&\multicolumn{2}{l|}{$^{~0.41}_{~0.25}$} \\
     & $2_3^+$ & $2_1^+$ & $0$&$610$
      & \multicolumn{4}{l|}{$~~>~0.16$} \\
     & $2_1^+$ & $2_1^+$ & $0$&$666^{*}$
      & \multicolumn{4}{c|}{---} \\
   \hline
     $^{54}$Cr & $2_2^+$ & $2_1^+$ & $0$&$021$
      & $0$&$023$&$~0$&$009$ \\
     & $2_3^+$ & $2_1^+$ & $0$&$519$
      & \multicolumn{4}{l|}{$~~>~0.10$} \\
     & $2_1^+$ & $2_1^+$ & $1$&$467^{*}$
      & $1$&$12$&$0$&$20^{*}$ \\
   \hline
     $^{56}$Fe & $2_2^+$ & $2_1^+$ & $0$&$476$
      & $0$&$233$&$0$&$072$ \\
     & $2_3^+$ & $2_1^+$ & $0$&$0005$ & $0$&$070$&$0$&$009$ \\
     & $1_1^+$ & $0_1^+$ & $0$&$0004$
      & \multicolumn{4}{l|}{$~~<~0.00001$} \\
     & $1_1^+$ & $2_1^+$ & $0$&$090$ & $0$&$040$&$0$&$003$ \\
     & $4_2^+$ & $4_1^+$ & $0$&$841$ & $0$&$206$&$0$&$038$ \\
     & $2_4^+$ & $2_1^+$ & $0$&$202$ & $0$&$109$&$0$&$045$ \\
     & $3_1^+$ & $2_1^+$ & $0$&$028$ & $0$&$054$&$0$&$013$ \\
     & $3_1^+$ & $4_1^+$ & $0$&$030$ & $0$&$098$&$0$&$023$ \\
     & $3_1^+$ & $2_2^+$ & $0$&$026$ & $0$&$018$&$0$&$007$ \\
   \hline
\end{tabular}
\begin{tabular}{|c||c|c||r@{.}l|r@{.}l@{$\pm$}r@{.}l|}
   \hline
     nucl. & $i$ & $f$ & \multicolumn{2}{c|}{Cal.}
      &  \multicolumn{4}{c|}{Exp.} \\
   \hline
     $^{56}$Fe & $1_2^+$ & $0_1^+$ & $0$&$061$
      & $0$&$047$&$0$&$013$ \\
     & $1_2^+$ & $2_1^+$ & $0$&$048$ & $0$&$054$&$0$&$014$ \\
     & $1_3^+$ & $0_1^+$ & $0$&$151$ & \multicolumn{4}{c|}{---} \\
     & $1_3^+$ & $2_1^+$ & $0$&$155$ & \multicolumn{4}{c|}{---} \\
     & $2_5^+$ & $2_1^+$ & $0$&$012$ & $0$&$003$&$0$&$002$ \\
     & $6_2^+$ & $6_1^+$ & $0$&$274$ & $1$&$09$&$0$&$18$ \\
     & $2_1^+$ & $2_1^+$ & $1$&$379^{*}$ & $1$&$20$&$0$&$20^{*}$ \\
   \hline
     $^{58}$Ni & $2_2^+$ & $2_1^+$ & $0$&$156$
      & $0$&$018$&$0$&$009$ \\
     & $1_1^+$ & $0_1^+$ & $0$&$00004$ & $0$&$0014$&$0$&$0003$ \\
     & $0_2^+$ & $1_1^+$ & $0$&$0001$ & \multicolumn{4}{c|}{---} \\
     & $0_3^+$ & $1_1^+$ & $0$&$0005$ & $0$&$158$&$0$&$009$ \\
     & $2_3^+$ & $2_1^+$ & $0$&$018$ & $0$&$134$&$0$&$021$ \\
     & $2_3^+$ & $2_2^+$ & $0$&$140$ & $0$&$555$&$0$&$090$ \\
     & $2_4^+$ & $2_1^+$ & $0$&$030$ & $0$&$070$&$0$&$029$ \\
     & $3_1^+$ & $4_1^+$ & $0$&$308$ & $0$&$16$&$0$&$14$ \\
     & $2_1^+$ & $2_1^+$ & $-0$&$168^{*}$
      & \multicolumn{4}{c|}{---} \\
   \hline
\end{tabular}
\\
$^*$) Magnetic moment.
\end{small}
\end{table*}

\clearpage
\subsection{$Z={\em odd},~N=30$ nuclei}
\label{subsec:M1_Zodd-N30}

The M1 properties are shown in Table~\ref{tab:M1_ZoN30},
for $Z={\em odd}$, $N=30$ nuclei.

The M1 quantities in $^{55}$Mn are reproduced remarkably well.
Together with the agreement in the E2 properties,
this fact suggests good convergence
of the wavefunctions in this nucleus.
The agreement in $^{57}$Co is only for the orders of magnitude.

\begin{table*}
\centering
\caption{\label{tab:M1_ZoN30}}
\begin{small}
$B({\rm M1})$ values ($\mu_{\rm N}^2$) or M1 static moments
($\mu_{\rm N}$) in $Z={\em odd}$, $N=30$ nuclei.
The experimental data are taken
from Refs.\cite{ref:NDS51,ref:NDS53,ref:NDS55,ref:NDS57}.\\
\begin{tabular}{|c||c|c||r@{.}l|r@{.}l@{$\pm$}r@{.}l|}
   \hline
     nucl. & $i$ & $f$ & \multicolumn{2}{c|}{Cal.}
      & \multicolumn{4}{c|}{Exp.} \\
   \hline
     $^{51}$Sc & $[{7\over 2}]_1^-$ & $[{7\over 2}]_1^-$
      & $5$&$061^{*}$ & \multicolumn{4}{c|}{---} \\
   \hline
     $^{53}$V & $[{5\over 2}]_1^-$ & $[{7\over 2}]_1^-$
      & $0$&$065$ & \multicolumn{4}{l|}{$~~>~0.027$} \\
     & $[{3\over 2}]_1^-$ & $[{5\over 2}]_1^-$
      & $0$&$00001$ & $0$&$0032$&$~0$&$0003$ \\
     & $[{7\over 2}]_1^-$ & $[{7\over 2}]_1^-$
      & $4$&$670^{*}$ & \multicolumn{4}{c|}{---} \\
   \hline
     $^{55}$Mn & $[{7\over 2}]_1^-$ & $[{5\over 2}]_1^-$
      & $0$&$042$ & $0$&$076$&$0$&$003$ \\
     & $[{9\over 2}]_1^-$ & $[{7\over 2}]_1^-$
      & $0$&$085$ & $0$&$197$&$0$&$021$ \\
     & $[{11\over 2}]_1^-$ & $[{9\over 2}]_1^-$
      & $0$&$227$ & $0$&$224$&$0$&$023$ \\
     & $[{3\over 2}]_1^-$ & $[{5\over 2}]_1^-$
      & $0$&$147$ & $0$&$167$&$0$&$036$ \\
     & $[{13\over 2}]_1^-$ & $[{11\over 2}]_1^-$
      & $0$&$154$ & $0$&$233$&$0$&$054$ \\
     & $[{15\over 2}]_1^-$ & $[{13\over 2}]_1^-$
      & $0$&$257$ & $0$&$251$&$0$&$072$ \\
     & $[{5\over 2}]_1^-$ & $[{5\over 2}]_1^-$
      & $3$&$378^{*}$ & $3$&$453$&$0$&$001^{*}$ \\
     & $[{7\over 2}]_1^-$ & $[{7\over 2}]_1^-$
      & $4$&$477^{*}$ & $4$&$4$&$0$&$7~^{*}$ \\
   \hline
     $^{57}$Co & $[{9\over 2}]_1^-$ & $[{7\over 2}]_1^-$
      & $0$&$304$ & $0$&$372$&$0$&$036$ \\
     & $[{1\over 2}]_1^-$ & $[{3\over 2}]_1^-$
      & $0$&$652$ & $0$&$090$&$0$&$011$ \\
     & $[{11\over 2}]_1^-$ & $[{9\over 2}]_1^-$
      & $0$&$722$ & $0$&$877$&$0$&$090$ \\
   \hline
\end{tabular}
\begin{tabular}{|c||c|c||r@{.}l|r@{.}l@{$\pm$}r@{.}l|}
   \hline
     nucl. & $i$ & $f$ & \multicolumn{2}{c|}{Cal.}
      &  \multicolumn{4}{c|}{Exp.} \\
   \hline
     $^{57}$Co & $[{7\over 2}]_2^-$ & $[{7\over 2}]_1^-$
      & $0$&$002$ & $0$&$022$&$0$&$003$ \\
     & $[{7\over 2}]_2^-$ & $[{9\over 2}]_1^-$
      & $0$&$115$ & $0$&$662$&$0$&$090$ \\
     & $[{5\over 2}]_1^-$ & $[{7\over 2}]_1^-$
      & $0$&$044$ & $0$&$236$&$0$&$032$ \\
     & $[{5\over 2}]_2^-$ & $[{7\over 2}]_1^-$
      & $0$&$002$ & \multicolumn{4}{c|}{---} \\
     & $[{7\over 2}]_3^-$ & $[{7\over 2}]_1^-$
      & $0$&$032$ & $0$&$003$&$0$&$001$ \\
     & $[{7\over 2}]_3^-$ & $[{9\over 2}]_1^-$
      & $0$&$775$ & $0$&$100$&$0$&$020$ \\
     & $[{9\over 2}]_2^-$ & $[{7\over 2}]_1^-$
      & $0$&$020$ & \multicolumn{4}{l|}{$~~<~0.034$} \\
     & $[{9\over 2}]_2^-$ & $[{9\over 2}]_1^-$
      & $0$&$028$ & \multicolumn{4}{l|}{$~~<~0.054$} \\
     & $[{9\over 2}]_2^-$ & $[{11\over 2}]_1^-$
      & $0$&$001$ & \multicolumn{4}{l|}{$~~<~0.21$} \\
     & $[{9\over 2}]_2^-$ & $[{7\over 2}]_2^-$
      & $0$&$004$ & $0$&$36$&$0$&$13$ \\
     & $[{13\over 2}]_1^-$ & $[{11\over 2}]_1^-$
      & $0$&$718$ & $0$&$73$&$0$&$18$ \\
     & $[{15\over 2}]_1^-$ & $[{13\over 2}]_1^-$
      & $0$&$730$ & $0$&$233$&$0$&$090$ \\
     & $[{7\over 2}]_1^-$ & $[{7\over 2}]_1^-$
      & $4$&$729^{*}$ & $4$&$727$&$0$&$009^{*}$ \\
     & $[{3\over 2}]_1^-$ & $[{3\over 2}]_1^-$
      & $2$&$353^{*}$ & $3$&$0$&$0$&$6~^{*}$ \\
   \hline
\end{tabular}
\\
$^*$) Magnetic moment.
\end{small}
\end{table*}

\clearpage
\subsection{Summary of electromagnetic properties}
\label{subsec:sum_EM}

We summarize the consequences for the electromagnetic properties.
The E2 transition probabilities and moments
are comprehensively reproduced,
together with the effective charges derived
by the Sagawa-Brown method.
The good agreement with the measurement
in the collective ({\it i.e.}, enhanced) E2 transitions
suggests that the convergence of the wavefunctions is reasonable
in the framework of the $k\leq 2$ truncation.
We also acquire an overall reproduction
of the M1 transition rates and moments,
by using Towner's single-particle $g$-factors.
Particularly, most of the measured magnetic moments
are precisely reproduced.
The remaining discrepancies may be ascribed
to the $k>2$ configuration or the sd-shell contribution.

\section{Electron scattering form factors}
\label{sec:FF}

We calculate longitudinal form factors
of inelastic electron scattering
with an angular momentum transfer of two units (C2),
from the ground state to several $2^+$ states
for a few even-even nuclei.
The form factors provide us with richer information
than the $\gamma$-transition probabilities,
owing to the off-shell photon exchange.
We here attempt to see
how well the present calculation reproduces
the quadrupole collective feature.
The method of Sagawa-Brown\cite{ref:Sagawa-Brown},
which has already been sketched in Section~\ref{sec:E2},
is used again.
This method was shown to work very well
for longitudinal form factors with collectivity
for nuclei around the doubly magic core.

The C2 form factors are calculated
from the shell model density matrices,
together with the renormalized single-particle transition densities
of Eq.(\ref{eqB:TD}).
The nucleon finite-size effect is incorporated
in the dipole approximation\cite{ref:BM1},
and the contribution of the center-of-mass motion is subtracted
in the harmonic oscillator approximation\cite{ref:cm-corr}.
The plane-wave Born approximation (PWBA) is employed,
taking into account the Coulomb distortion effect
in terms of the effective momentum transfer $q_{\rm eff}$.
It should be emphasized that this calculation
includes no additional adjustable parameters,
as pointed out in Section~\ref{sec:E2}.

The results for $^{56}$Fe were already shown in Ref.\cite{ref:Nakada}.
Here we mention that the results of this calculation
agree with the data\cite{ref:ee'Fe56} very well.
Particularly, the form factor in the excitation to $2_1^+$
is in a precise agreement.
It is also remarkable that the collectivity of the $2_4^+$
is correctly reproduced.

The C2 form factors in $^{54}$Cr are shown in Fig.\ref{fig:C2Cr54}.
In the experiments
the transverse mode was not separated,
though most of the reaction is known
to be dominated by the C2 mode\cite{ref:deForest-Walecka}.
The experimental data are described precisely
with respect to the $2_1^+$ state.
As for the $2_2^+$ state, the shape is reproduced well,
though the absolute value is overall underestimated
by a factor of $2$.
It is pointed out that this transition is not collective.
There are no reported data
with respect to the $2_3^+$ and $2_4^+$ states.
We show them, however,
since they are good candidates
for the mixed-symmetry state\cite{ref:Lieb}.

\begin{figure}
\centering
%\vspace{60mm}
\vspace{30mm}
\caption{\label{fig:C2Cr54}}
\begin{small}
$(e,e')$ form factors from the ground state
to the lowest four $2^+$ states of $^{54}$Cr.
The solid lines show the shell model results
with the core polarization effect evaluated by the HF+RPA.
Those without core polarization are shown by dotted lines.
The crosses exhibit the experimental data
taken from Ref.\cite{ref:Lightbody}.
The transverse mode is not separated in the experiment.
\end{small}
\end{figure}

The form factors from $0_1^+$ to $2_1^+$
in $^{54}$Fe and $^{58}$Ni are shown
in Fig.\ref{fig:C2Fe54} and Fig.\ref{fig:C2Ni58}, respectively.
The absolute value is underestimated,
and this is consistent with the $B({\rm E2})$ values
discussed in Section~\ref{sec:E2}.
This might indicate a worse convergence of the wavefunctions
for these nuclei.
Nevertheless, the shape of the form factors,
especially the position of the peaks and the dips
are reproduced very well.

\begin{figure}
\centering
%\vspace{40mm}
\vspace{30mm}
\caption{\label{fig:C2Fe54}}
\begin{small}
$(e,e')$ form factor from the ground state
to the $2_1^+$ state of $^{54}$Fe.
See the caption of Fig.\ref{fig:C2Cr54}.
The experimental data are taken from Ref.\cite{ref:ee'Fe54}.
\end{small}
\end{figure}

\begin{figure}
\centering
%\vspace{40mm}
\vspace{30mm}
\caption{\label{fig:C2Ni58}}
\begin{small}
$(e,e')$ form factor from the ground state
to the $2_1^+$ state of $^{58}$Ni.
See the caption of Fig.\ref{fig:C2Cr54}.
The experimental data are taken from Ref.\cite{ref:ee'Ni58}.
\end{small}
\end{figure}

\section{Proton scattering cross sections}
\label{sec:pp'}

Recently $(\vec p, p')$ experiments are performed
for $^{56}$Fe and $^{54}$Cr\cite{ref:Takamatsu},
with an incident proton energy of 65MeV.
The differential cross sections and the analyzing powers
are extracted.
The data are analyzed in the distorted-wave Born approximation (DWBA),
by using the present shell model density matrices.
This DWBA calculation is carried out
by Takamatsu and his collaborators\cite{ref:Takamatsu}.
We briefly review their results in this section.

The Bonn-J\"{u}lich effective interaction\cite{ref:Bonn-Juelich}
is employed as the interaction
between the incident or scattered proton
and the nucleons in the target nucleus.
Although the M3Y interaction\cite{ref:M3Y}
and the Paris-Hamburg interaction\cite{ref:Paris-Hamburg}
are tried also,
the results were essentially the same.
In order to reproduce the experimental data,
an overall normalization factor is used for each transition.
This factor would correspond to the core polarization effect.
In fact, the adjusted normalization factors are consistent
with the ones expected from the effective charges
employed in the description of the $B({\rm E2})$ values.

The absolute value of the cross section
in the excitation to the $2_3^+$ state
is smaller by an order of magnitude
than to the $2_2^+$ or the $2_4^+$ states
in the experimental data of $^{56}$Fe.
This is consistent with the present calculation.
In addition, the excitation to the $2_3^+$ state
shows a strikingly anomalous angular distribution.
The present shell model wavefunctions reproduce
this anomaly remarkably well.

The experimental differential cross sections
are also excellently reproduced for $^{54}$Cr\cite{ref:Takamatsu}.

We have not seen any notable contradiction
between experiments and the present calculation
with respect to $(e,e')$ and $(p,p')$ data.

\section{Summary}
\label{sec:summary}

The properties of the low-lying states
in $20<Z\leq 28$, $28\leq N\leq 30$ nuclei
have been investigated from a microscopic standpoint.
The model space has been restricted to the pf-shell,
while we have incorporated the excitation
from ${\rm 0f}_{7/2}$
to $({\rm 0f}_{5/2} {\rm 1p}_{3/2} {\rm 1p}_{1/2})$
up to two particles.
This space leads to one of the largest-scale shell model calculations
that have ever been carried out.

We have adopted the Kuo-Brown interaction
on top of the $^{40}$Ca core,
including the $3p$-$1h$ correction.
This effective interaction is derived
from the Hamada-Johnston $NN$ potential,
through the $G$-matrix calculation.
Despite the failure near the beginning of the pf-shell,
it has been shown for the first time, by this work,
that this microscopic interaction produces
an excellent description of the nuclei
around the middle of the pf-shell.

The observed energy levels are reproduced
by the present shell model calculation in a wide range of energy,
for any of the $20<Z\leq 28$, $28\leq N\leq 30$ nuclei.
The energy range of good agreement is typically
$Ex\lsim 4$MeV for even-even nuclei,
$Ex\lsim 2.5$MeV for odd-mass nuclei
and $Ex\lsim 2$MeV for odd-odd nuclei,
with the discrepancies of $\delta E \lsim 0.3$MeV.
These energy ranges are notably wider
than those obtained in previous shell model
studies\cite{ref:Ogawa,ref:Ogawa2,ref:Raz-etal,ref:Motoba-etal},
covering the region where candidates for the mixed-symmetry states
were reported.
Among a number of states dominated by the $k=0$ configuration,
more than twenty states in the low energy region
are found to have $k=1$ probabilities larger than $k=0$ ones.
The $k=1$ dominance of those states is consistent with the fact
that the Horie-Ogawa calculation is not capable of describing them.
Those states are also reproduced as well as the $k=0$ dominant ones.
The present calculation is undoubtedly more successful
than any other calculation that has ever been reported
in this mass region.

In order to study the E2 properties,
we have calculated the effective charges
by applying Sagawa and Brown's HF+RPA method.
The measured E2 moments and transition rates
are reproduced fairly well,
by the shell model wavefunctions
together with the microscopic effective charges.
The M1 properties have also been investigated
by using Towner's microscopic parameters.
This approach gives an overall agreement with experiment.
There remains, however, some discrepancies
in the electromagnetic properties.
Most of the discrepancies seem to be
due to the influence of the $k>2$ or the sd-shell configurations.
The overall reproduction of the electromagnetic properties
has confirmed the reliability of the shell model wavefunctions.

The $(e,e')$ C2 form factors have also been calculated
in the same framework as the E2 properties.
Together with the $(p,p')$ results,
we have confirmed that the quadrupole collective features
are successfully described by the present shell model wavefunctions.

We have investigated the low-lying states of the nuclei
in the middle of the pf-shell, from a microscopic standpoint.
The present effective interaction is derived on a realistic basis,
without fitting two-body matrix elements
to observed energy levels,
although there remains an uncertainty
in the $G$-matrix calculation.
By applying this interaction, some of real situations
of the low-lying states of the middle pf-shell nuclei
are successfully reproduced.
It is remarked that the large-scale calculation
including the excitation from ${\rm 0f}_{7/2}$
to $({\rm 0f}_{5/2} {\rm 1p}_{3/2} {\rm 1p}_{1/2})$
up to two nucleons is crucial for this agreement.
The successful results of this study enable us to place confidence,
to a certain extent,
in the ways of understanding the nuclear many-body system
from the nucleonic degrees of freedom.

$\\$
\noindent
The authors are grateful to Prof. H. Sagawa
for valuable discussion of the HF+RPA calculation.
The authors thank Prof. A. Gelberg for careful reading the manuscript.
The computers used for the numerical calculations are
HITAC-S820/80 in Computer Center, University of Tokyo,
HITAC-S820/80 in Kanagawa Factory, Hitachi Corporation,
and VAX6440 in Meson Science Laboratory, University of Tokyo.
This work is financially supported, in part,
by Research Center for Nuclear Physics, Osaka University.

\end{document}